\documentclass[tradiabstract]{aa_hack}
\usepackage{hyperref}
\usepackage{graphicx}
\usepackage{amsbsy}
\usepackage{amsmath}
\usepackage{natbib}
\usepackage{color}
\usepackage{xcolor}
\usepackage{txfonts}
\usepackage{url}
\usepackage{bm}

\newcommand{\percent}[1]{\%}
\DeclareMathSymbol{\varOmega}{\mathord}{letters}{"0A}
\DeclareMathSymbol{\varSigma}{\mathord}{letters}{"06}
\DeclareMathSymbol{\varPsi}{\mathord}{letters}{"09}
\DeclareMathSymbol{\varPhi}{\mathord}{letters}{"08}
\DeclareMathSymbol{\varGamma}{\mathord}{letters}{"00}

\begin{document}

\title{Self-oxidation of the atmospheres of rocky planets\\ with implications
for the origin of life}
\titlerunning{Self-oxidation of the atmospheres of rocky planets}

\author{Anders Johansen$^{1,2}$, Eloi Camprubi$^3$, Elishevah van Kooten$^{1}$,
H.~Jens Hoeijmakers$^2$}
\authorrunning{Johansen et al.}

\institute{$^1$ Center for Star and Planet Formation, Globe Institute,
University of Copenhagen, \O ster Voldgade 5-7, 1350 Copenhagen, Denmark \\ $^2$
Lund Observatory, Department of Physics, Lund University, Box 43, 221 00 Lund,
Sweden \\ $^3$ School of Integrative Biological and Chemical Sciences,
University of Texas Rio Grande Valley, 1201 W University Dr, Edinburg TX 78539,
USA \url{Anders.Johansen@sund.ku.dk}}

\date{}

\abstract{Rocky planets may acquire a primordial atmosphere by
outgassing of volatiles from their magma ocean. The distribution of O
between H$_2$O, CO and CO$_2$ in chemical equilibrium subsequently changes
significantly with decreasing temperature. We explore here two chemical models:
one where CH$_4$ and NH$_3$ are assumed to be irrevocably destroyed by
photolysis, and one where these molecules persist. In the first case, we show
that CO cannot co-exist with H$_2$O, since CO oxidizes at low temperatures to
form CO$_2$ and H$_2$. In both cases, H escapes from the thermosphere within a
few ten million years by absorption of stellar XUV radiation. This escape drives
an atmospheric self-oxidation process whereby rocky planet atmospheres become
dominated by CO$_2$ and H$_2$O, regardless of their initial oxidation state at
outgassing.  HCN is considered a potential precursor of prebiotic compounds and
RNA. Our oxidizing atmospheres are inefficient at producing HCN by lightning.
Instead, we demonstrate that lightning-produced NO, which dissolves as nitrate
in the oceans, and interplanetary dust particles may be the main sources of
fixed nitrogen to emerging biospheres. Our results highlight the need for
origin-of-life scenarios where the first metabolism fixes its C from CO$_2$,
rather than from HCN and CO.}

\keywords{Earth -- meteorites, meteors, meteoroids -- planets and satellites:
formation -- planets and satellites: atmospheres -- planets and satellites:
composition -- planets and satellites: terrestrial planets}

\maketitle

\section{Introduction}

Understanding the prevalence of life in our Galaxy is an inherently
multidisciplinary endeavour that binds together research efforts from a large
range of scientific fields. Astronomical observations have detected now a
significant number of small rocky planets beyond the Solar System
\citep{Kossakowski+etal2023}, with successful characterization so far of their
orbits, masses and/or radii as well as the effective temperature
\citep[e.g.][]{Anglada-Escude+etal2016,Gillon+etal2017}. The imminent
characterization of the atmospheric pressure and composition will be key to
assessing the potential habitability of such rocky exoplanets. Recently, thermal
flux measurements with the flagship James Webb Space Telescope have provided
evidence for the lack of atmospheres around the rocky planets TRAPPIST-1b
(effective temperature 400 K) and for either a thin atmosphere or lack of an
atmosphere on TRAPPIST-1c (effective temperature 340 K)
\citep{Greene+etal2023,Zieba+etal2023}. With the first characterization of an
atmosphere of a rocky exoplanet within reach, this emphasizes a fundamental
research question: what is the expected composition of the secondary atmosphere
outgassed from the magma ocean of a rocky planet after its formation? And how
does that composition reflect on possible scenarios for the origin of life on
Earth and on other planets?

Many early Earth environments have been proposed as the cradle of life on our
planet. These environments range from submarine hydrothermal systems
\citep{RussellHall1997,Sojo+etal2016} to surficial hydrothermal geysers
\citep{DamerDeamer2020} and volcanic pools \citep{FoxStrasdeit2013}, and also
include more niche environments such as floating pumice rafts
\citep{Brasier+etal2011}, nuclear-powered geysers \citep{EbisuzakiMaruyama2017},
mica clay sheets \citep{Hansma2013}, and hydrothermal-sedimentary settings
\citep{Westall+etal2018}. Each of these environments have
their strong and weak points when it comes to promoting or hampering certain
types of prebiotic chemistries which are universally deemed as important
(e.g.~nitrogen and carbon fixation, C-N bond formation, C-C bond formation,
condensation/polymerization reactions). We will not delve into deeper
discussions of the pros and cons for each locale and hypothesis here. Instead,
we will contextualize our findings on the oxidation state of primitive
atmospheres with the associated types of prebiotic chemistry that in turn become
more or less feasible as a consequence.

Regardless of the primordial environment in which organic molecules (i.e.,
molecules hosting carbon bonded to other carbon, nitrogen or hydrogen atoms)
started self-organizing and displaying life-like behaviors, it is important to
address whether the environmental feedstock molecules needed to form such
compounds were synthesized within Earth (endogenous source) or were delivered
from extraterrestrial (exogenous) sources -- or a mixture of both. It can be
argued that the relevance of exogenous delivery as a prebiotic chemistry
promoter relies not so much on the type of organics delivered, but in the sheer
quantity of fixed carbon and nitrogen being delivered versus that which is
endogenously synthesized. However, recent research points to a nuanced
cross-play between many prebiotic molecules and their complex network chemistry
\citep{XavierKauffman2022}, suggesting that the type of organics – and not only
their quantities – may have had an important impact on the pathway to
abiogenesis (the emergence of life from non-living matter). 

There is currently no consensus on which organic (as well as inorganic) species
are prebiotically relevant as feedstock molecules. Some authors prefer to use
ancient life as a guide, so that the jump between prebiotic chemistry and the
universally conserved core of biochemistry is as short as possible
\citep[e.g.][]{HarrisonLane2018}.  This school of thought explores the
non-enzymatic chemistry between molecules and reaction types which are directly
represented in (or are very close to) the chemistry of current biomolecules and
their biological precursors – for example, using N$_2$/NH$_3$ and CO$_2$ for the
synthesis of biomolecules containing both C and N atoms, such as amino acids and
nucleotides. Alternatively, the main competing school of thought favors the
usage of more exotic (to life-as-we-know-it) chemical precursors such as HCN and
its derivatives in a highly reducing atmosphere
\citep[e.g.][]{Haldane1929,Oparin1938,WuSutherland2019}, since this type of
simple C-N-bearing molecules are very reactive and thus readily build up larger
biomolecules such as amino acids and nucleotides. This type of chemistry bares a
strong parallelism to that explored in the famous \cite{MillerUrey1959}
experiment, which considered lightning in a heavily reducing atmosphere. Such a
scenario must then postulate that once biomolecules are obtained and the system
eventually becomes a living one, Darwinian evolution will – over generations –
radically modulate the original ‘alien’ prebiotic chemistry towards the
CO$_2$-based metabolism used by the Last Universal Common Ancestor \citep[LUCA,
see][]{BoydPeters2013}.  The prebiotic plausibility of these two approaches
relies heavily on the chemical composition (and thus the overall redox state) of
the primitive atmosphere. More neutral or mildly oxidizing atmospheres
(dominated by CO$_2$) will naturally promote prebiotic chemistry that utilizes
CO$_2$ and N$_2$, whereas a reducing atmosphere (dominated by CO, CH$_4$, and
H$_2$) will much more easily promote HCN-related reactivity.

In this paper we focus therefore on first-principles modelling of the potential
composition of the first atmosphere of Earth and other rocky planets. We are
particularly interested in understanding the oxidation state of the atmosphere.
The oxidation state is maybe the most important unknown in origin-of-life
discussions, since the very nature of equilibrium chemistry changes
fundamentally between slightly reducing atmospheres where H$_2$, CO, CH$_4$
and NH$_3$ can exist in significant amounts to oxidizing atmospheres consisting
almost entirely of CO$_2$, H$_2$O and N$_2$.

The paper is organized as follows. In Section 2 we present our chemical
equilibrium model for the primordial atmosphere and elucidate the role of
atmospheric mass loss in oxidizing initially reducing atmospheres. In Section 3
we discuss fixation of nitrogen in atmospheres of various degrees of oxidation
and confirm the known result that N$_2$ is fixed by lightning to HCN in reducing
atmospheres and to NO in oxidizing atmospheres. We show that the internal
production of NO in atmospheres with a realistic oxidation state provides an
emerging biosphere with comparable amounts of usable nitrogen relative to
external delivery by interplanetary dust particles. In Section 4 we discuss the
implications of our model for the origin of life; particularly we emphasize that
we expect that rocky planet atmospheres will either be oxidizing at the
magma ocean outgassing stage or undergo later self-oxidation by escape of the H
component by XUV irradiation. We conclude briefly on our results in Section 5.
Appendix A describes our XUV mass loss model and discusses various
parameter variations to the nominal model. Appendix B contains a brief estimate
of the possible contribution to fixed N from early asteroid impacts.

\section{Atmosphere equilibrium model}
\label{s:equilibrium}

The differentiated core-mantle structure of the terrestrial planets in the Solar
System provides direct evidence that these objects reached the melting
temperature of metal and likely also silicate rock during their formation.
The formation channel of Earth is debated, with classical giant impact models
yielding formation time-scale of up to 100 million years
\citep{RaymondMorbidelli2022}. In contrast, the pebble accretion model instead
postulates Earth formation within the few-million-years life-time of the gaseous
protoplanetary disc \citep{Johansen+etal2021}, which must be supplemented by a
single moon-forming giant impact after at least 35 million years later to comply
with the low amount of $^{182}$W in the Earth's mantle
\citep{YuJacobsen2011,Johansen+etal2023b,OlsonSharp2023}.  Mars, in contrast,
has been inferred from the same Hf-W system to have formed its core within just
a few million years after the formation of the Sun \citep{DauphasPourmand2011}.
The immense rate of accretion energy released in the compilation of such rocky
bodies must under all circumstances lead to their partial or full melting and
prevalence of global magma oceans on newborn rocky planets
\citep{MatsuiAbe1986,Elkins-Tanton2008,Elkins-Tanton2012}.

We assume that the atmosphere over the magma ocean evolves in chemical
equilibrium with volatiles dissolved in the underlying magma
\citep{Elkins-Tanton2008,Sossi+etal2020,Johansen+etal2023c}. These volatiles are
delivered to a growing planet by both planetesimals and pebbles. Enstatite
chondrites and ordinary chondrites (the meteorite classes representing the
likely composition of planetesimals that formed in the 1 AU region around the
Sun) contain significant amounts of hydrogen, carbon and nitrogen
\citep{Grewal+etal2019,Piani+etal2020}, although measurements of a high hydrogen
abundances in enstatite chondrites were critized recently by
\cite{Peterson+etal2023}. \cite{Johansen+etal2021} nevertheless demonstrated
that these volatiles can also be delivered during the protoplanetary disc phase
via rims of ice and organics on small pebbles.  Ambient conditions above
$\approx$400 K within the protoplanetary disc would lead to sublimation of water
and destruction of most organic carriers of C and N \citep{GailTrieloff2017}.
Formation of rocky planets under such hot conditions could lead to a largely
volatile-free magma ocean, as was likely the case for Mercury. Mercury's mantle
indeed has a very low fraction of oxidized iron
\citep{VanderKaadenMcCubbin2015}, in agreement with the picture that Mercury
provides an example from the Solar System of a planet formed well interior of
the water ice-line. Such a planet could later obtain an atmosphere by delivery
of volatiles via impacts of icy asteroids and comets that formed under much
colder conditions. We focus instead this work on studying rocky planet
atmospheres that are outgassed from the magma ocean, under the assumption that
this was the case for Venus, Earth and Mars and that magma ocean degassing
provides the dominant pathway for rocky exoplanets to acquire their most
primordial atmosphere.

We set up here an atmosphere model comprising four basic volatile-building
elements: H, C, N and O. We assume that the crystallisation of the magma ocean
occurs from the bottom and up towards the surface (the crystallisation takes
place over 0.01--1 Myr years after the protoplanetary disc dissipates and
planetary accretion terminates\footnote{The time-scale depends sensitively on
the assumed abundances and opacities of H$_2$O and CO$_2$; see
\cite{Zahnle+etal1988}, \cite{Elkins-Tanton2008} and
\cite{Johansen+etal2023b}.}). This leads to efficient degassing of the volatiles
from the magma and fixes the respective numbers of H, C, N and O in the
atmosphere, with only minor H, C and N persisting in the solid mantle but with
major amounts of mantle oxygen remaining bound to iron and silicates. The
molecular speciation will nevertheless continue to undergo changes as the
temperature sinks towards its final equilibrium with stellar incoming radiation.
We analyse this equilibrium speciation in Sections 2.1--2.4 for a range of
possible atmospheric oxidation states from strongly oxidizing to strongly
reducing.

We assume that H, C and N dissolve in the magma ocean in concentrations that are
in equilibrium both with partitioning to the core and with the partial pressures
of their molecular hosts in the atmosphere
\citep{Speelmanns+etal2019,Li+etal2020,Fischer+etal2020,Lichtenberg+etal2021,Johansen+etal2023b};
with the magma ocean composition assumed to be constant with depth due to mixing
by vigorous convection
\citep{Rubie+etal2003,Elkins-Tanton2008,Armstrong+etal2019}. While the magma
ocean is liquid, we further assume that the partial pressure of O$_2$ over
the magma ocean surface is set by buffer reactions in the magma between oxygen
and iron in its three oxidation states: Fe$^0$ (metallic Fe), Fe$^{2+}$ (FeO)
and Fe$^{3+}$ (FeO$_{1.5}$) \citep{Armstrong+etal2019,Deng+etal2020}. The oxygen
pressure in turn determines the speciation of H between H$_2$ and H$_2$O and of
C between CO and CO$_2$ \citep{Ortenzi+etal2020}. We therefore calculate
realistic atmospheric oxidation states as a function of the planetary mass in
Section 2.5.
\begin{figure*}
  \begin{center}
    \includegraphics[width=0.8\linewidth]{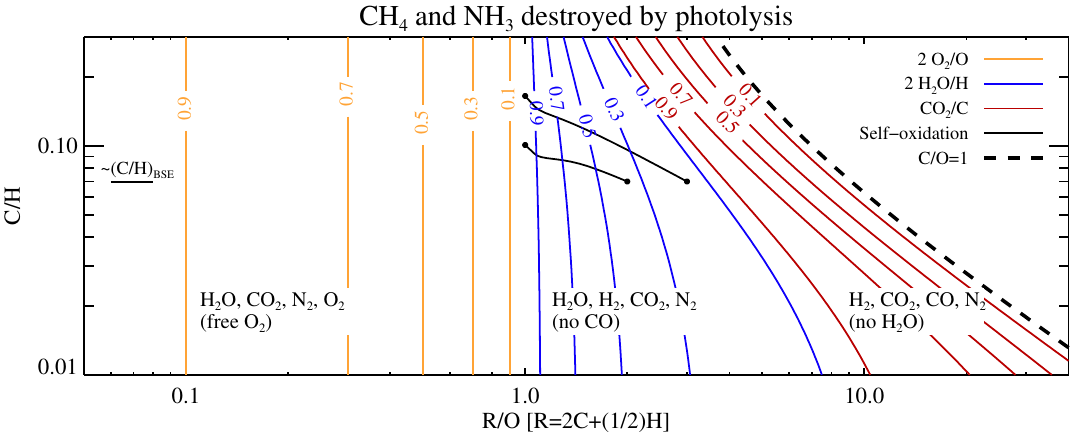}
    \includegraphics[width=0.8\linewidth]{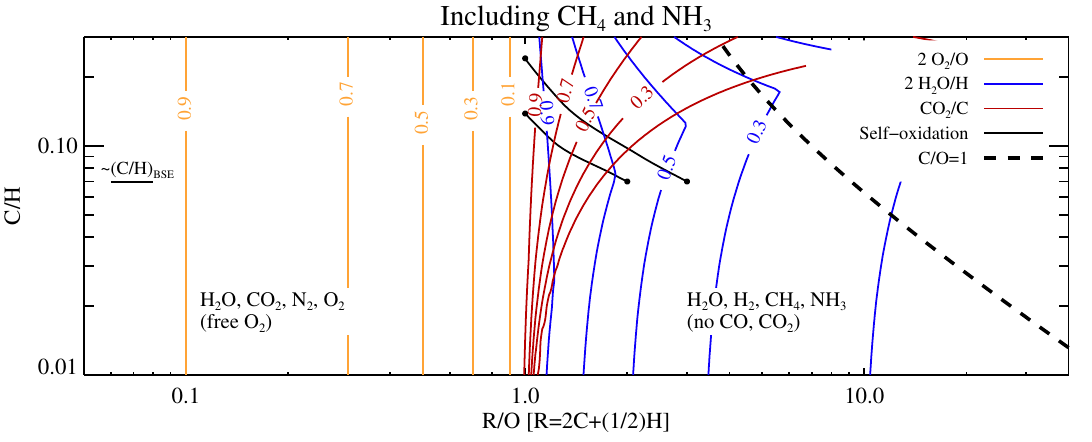}
  \end{center}
  \caption{The top panel shows, for a model where CH$_4$ and NH$_3$ are
  assumed to be destroyed by photolysis, the equilibrium abundances of O$_2$,
  H$_2$O and CO$_2$ at a total pressure of 100 bar as a function of the
  reduction parameter R/O and the C/H ratio, calculated at 500 K and normalized
  to their characteristic elemental constituent (O, H and C, respectively).  For
  ${\rm R/O}$$<$$1$, free oxygen must exist and O$_2$ even becomes the dominant
  O-carrier for very low values of R/O. At R/O$>$$1$, H$_2$O is first deprived
  of O and transformed to H$_2$, oxidizing in the process all CO molecules to
  CO$_2$. However, at large values of R/O, the O budget becomes insufficient to
  maintain 100\% CO$_2$; this leads to an increase in the fraction of C that
  resides in CO until the critical point of ${\rm C/O}$$=$$1$ is reached.  The
  black lines with dotted start and end points indicate self-oxidation tracks by
  H loss (from the data in Figure \ref{f:atmosphere_loss}, starting at
  R/O=2 and R/O=3 for the 50\% rotator host star and including the diffusion
  limit).  The bottom panel shows results when including CH$_4$ and NH$_3$ in
  the chemical equilibrium. These two molecules become the dominant C and N
  hosts for high R/O and low C/H. CO is nearly absent in the lower panel for the
  low C/H ratios tested here, which implies that the CH$_4$ abundance can be
  read off directly from the CO$_2$ curves through CH$_4$/C$\approx$1-CO$_2$/C.}
  \label{f:simple_atmosphere}
\end{figure*}

Young stars have X-ray luminosities up to 600 times the current solar X-ray
output and EUV luminosities up to 100 times the current value
\citep{Johnstone+etal2021}. These high-energy luminosities typically decay by an
order of magnitude over the first 100 million years of rotational slow-down.
Molecules and atoms have large cross sections to XUV absorption, leading to
heating of the thermosphere. As the lightest element, H escapes most easily when
heated and drags along the heavier atoms and molecules of the atmosphere
\citep{Sekiya+etal1980,ZahnleKasting1986,Erkaev+etal2014}. In Section 2.6 we
finally expose our atmospheres to mass loss by XUV irradiation and conclude that
loss of H drives an efficient self-oxidation of the atmosphere, regardless of
its initial oxidation state.

\subsection{Quantifying the oxidation state}

We choose to quantify the oxidation degree of the atmosphere through the ratio
R/O, with O denoting the number of oxygen atoms and R denoting a reducing
``pseudo-element'' that we introduce here and define as
\begin{equation}
  {\rm R} = 2 {\rm C} + (1/2) {\rm H} \, .
\end{equation}
For ${\rm R/O}$$<$$1$, the atmosphere must contain an excess of free O$_2$,
while ${\rm R/O}$$>$$1$ entails that there is not sufficient O to fully oxidize
H$_2$ to H$_2$O and C to CO$_2$. With additional knowledge of the bulk
atmospheric C/H ratio, the oxidation parameter R/O provides both the C/O and H/O
ratios through the relations
\begin{eqnarray}
  \frac{\rm C}{\rm O} &=& \frac{2 ({\rm R}/{\rm O}) ({\rm C}/{\rm H})}{1+4 ({\rm
  C}/{\rm H})} \, , \label{eq:COr} \\
  \frac{\rm H}{\rm O} &=& \frac{2 ({\rm R}/{\rm O})}{1+4 ({\rm C}/{\rm H})} \, .
  \label{eq:HO} 
\end{eqnarray}
Since most of the original carbon content of Earth's primordial atmosphere is
now stored in the mantle by subduction of carbonates
\citep{Sleep+etal2001,Kadoya+etal2020}, we calculate the probable C/H ratio of
the primordial bulk silicate Earth (which excludes any volatile reservoirs in
the core) by dividing the C contents of Venus' atmosphere with the H contents of
the Earth's oceans. Estimates of the C content buried in Earth's mantle are
comparable to the measured C contents of the Venus atmosphere
\citep{DasguptaHirschmann2010}, so we prefer to use the more direct measurement
from Venus rather than rely on indirect estimates. Our approach yields $({\rm
C/H})_{\rm BSE} \approx 0.07$ (by number). We emphasize that the majority of the
H and C budgets of the terrestrial planets likely reside in their cores
\citep{Li+etal2020,Fischer+etal2020,Johansen+etal2023c}, but we assume that
these reservoirs are not in contact with the magma ocean since it crystallizes
from the bottom up \citep{Elkins-Tanton2008} and are hence not relevant for
estimating the C/H ratio of the primordial atmosphere. This estimate of the
primordial C/H ratio also assumes that the Earth's surface reservoir of water
has not been depleted significantly with time by hydrogen escape, that it was
not changed by impacts with volatile-rich asteroids after the main accretion
phase and that only insignificant amounts of volatiles are trapped in the mantle
as the magma ocean crystallizes \citep[we refer to][for a perspective on this
assumption]{Hier-MajumderHirschmann2017}.

An additional assumption on the N/H ratio yields the atmospheric mixing
ratios of H, C, N and O through the expressions
\begin{eqnarray}
  X_{\rm H} &=& \frac{1}{1+{\rm C/H}+{\rm N/H}+{\rm O/H}} \, , \\
  X_{\rm C} &=& \frac{1}{{\rm H/C}+1+{\rm N/C}+{\rm O/C}} \, , \\
  X_{\rm N} &=& \frac{1}{{\rm H/N}+{\rm C/N}+1+{\rm O/N}} \, , \\
  X_{\rm O} &=& \frac{1}{{\rm H/O}+{\rm C/O}+{\rm N/O+1}} \, .
\end{eqnarray}
All quantities involved in evaluating the right-hand-sides of these equations
can be obtained from C/H, N/H, C/O and H/O, the latter two obtained from R/O
through equations (\ref{eq:COr}) and (\ref{eq:HO}). We set here ${\rm
N/H}=0.008$ by dividing the N contents of Venus' modern atmosphere with the H
contents of Earth's surface water reservoir.  Together with the total pressure,
the four quantities $X_{\rm H}$, $X_{\rm C}$, $X_{\rm N}$ and $X_{\rm O}$ now
completely describe the atmosphere after the crystallization of the magma ocean.
We do not consider here any further exchange of material between atmosphere and
mantle after the crystallization of the magma ocean.

\subsection{Equilibrium speciation}

In our nominal model, we calculate the equilibrium speciation of the four
basic elements into molecules using the simplified thermochemical model of
\cite{Ortenzi+etal2020} that conserves all four included elements. The chemical
equilibrium is given by balance of the two-way reaction set
\begin{eqnarray}
  {\rm H_2} + (1/2) \, {\rm O_2} &\rightleftharpoons& {\rm H_2O} \, ,
  \label{eq:H2} \\
  {\rm CO} + (1/2) \, {\rm O_2} &\rightleftharpoons& {\rm CO_2} \label{eq:CO} \,
  .
\end{eqnarray}
In this model, N is hosted entirely in the strongly bound N$_2$. We ignore
here any exchange of O between Fe$^{2+}$/Fe$^{3+}$ in the mantle and
H$_2$/H$_2$O in the atmosphere, under the assumption that the magma ocean
crystallizes rapidly after the end of the accretion \citep{KiteSchaefer2021} and
that Fe in the crust does not react with H$_2$O or O$_2$. At low temperatures
\citep[but not too low pressure, see][]{TianHeng2024}, two more reactions become
relevant for determining the main host of C and N \citep[see discussion
in][]{HengTsai2016},
\begin{eqnarray}
  {\rm CO} + 3 \, {\rm H_2} &\rightleftharpoons& {\rm CH_4} + {\rm H_2O} \, ,
  \label{eq:CH4} \\
  {\rm N_2} + 3 \, {\rm H_2} &\rightleftharpoons& 2 \, {\rm  NH_3}
  \label{eq:NH3} \, .
\end{eqnarray}
In oxidizing atmospheres, with R/O$<$$1$ and negligible free H$_2$, the
reactions with H$_2$ in equation (\ref{eq:CH4}) and (\ref{eq:NH3}) are
unimportant. However, when transitioning to reducing conditions with R/O$>$$1$,
the equilibrium C host molecule shifts from CO to CH$_4$ and the equilibrium N
host shifts from N$_2$ to NH$_3$ at the temperatures below 500--600 K relevant
for the early atmosphere \citep{Hirschmann2012,Sossi+etal2020}. Both CH$_4$ and
NH$_3$ nevertheless have large cross sections to destruction by UV photolysis
\citep{KuhnAtreya1979,Kasting1982,Romanzin+etal2005,Zahnle+etal2013}; NH$_3$ is
photolyzed at photon wavelengths between 160 nm and 230 nm and CH$_4$ below 145
nm \citep{Kasting2014}. \cite{Zahnle+etal2020} concluded that oxidation of
CH$_4$ to CO is the major destruction route of CH$_4$ in models of reducing
impact-produced atmospheres, particularly when the top of the atmosphere
contains significant amounts of water vapor. \cite{Line+etal2011} found
efficient photolysis of CH$_4$ at pressures below $10^{-5}$-$10^{-6}$ bar in
models of the hot Neptune GJ 436 b. \cite{Kasting2014} estimated the time-scale
to photolyse and oxidize 10 bar of CH$_4$ to scale as
\begin{equation}
  \tau_{\rm CH_4} = 30\,{\rm Myr} \left( \frac{P_{\rm CH_4}}{10\,{\rm bar}}
  \right) \left( \frac{F_{\rm UV}}{5 F_{\rm UV,\odot}} \right) \, .
  \label{eq:tCH4}
\end{equation}
This equation is normalized here, as in \cite{Kasting2014}, to a UV flux of 5
times the modern value, relevant for the 0.1 to 1 Gyr epoch of a relatively
slowly rotating star (see Figure \ref{f:XUV_luminosity_time}). In Section
\ref{s:Hloss} we discuss that during the earliest epoch of stellar evolution,
out to 100 Myr after planet formation, the XUV fluxes were 1-2 orders of
magnitude higher than that of the modern Sun. Therefore, the CH$_4$ photolysis
time-scale would likely have been $\sim$1 Myr during the crystallization of the
magma ocean and the cooling of the mantle. Importantly, the surface temperature
in our models is set to 500 K and hence reformation of CH$_4$ in the main
atmosphere is kinetically inhibited (see discussion in Section 2.4). Other
works \citep{Zahnle+etal2020,Wogan+etal2023} have reported longer life-times of
CH$_4$ compared to equation (\ref{eq:tCH4}). To bracket reality, we
therefore present in Section \ref{s:CH4NH3} the results of a model where
the abundances of both CH$_4$ and NH$_3$ are allowed to reach their equilibrium
values in the absence of photolysis\footnote{We note that even small amounts of
CH$_4$ in a reducing atmosphere can have relevance for origin-of-life scenarios,
due to its role in formation of HCN by photochemistry under mild UV irradiation
\citep{Pearce+etal2022}. We nevertheless focus here on exploring the two
end-member scenarios of either full destruction of CH$_4$ during atmosphere
cooling or full survival at the equilibrium concentration.}.
\begin{figure}
  \begin{center}
    \includegraphics[width=0.9\linewidth]{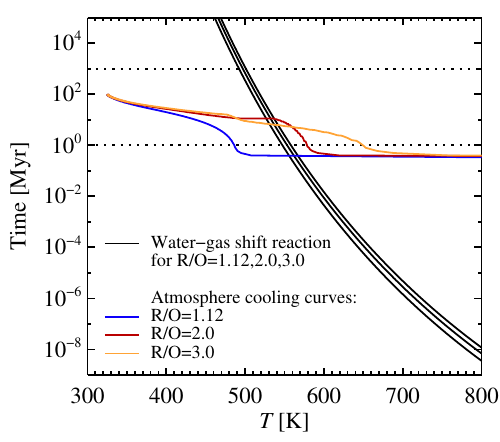}
  \end{center}
  \caption{The reaction time-scale for the water gas-shift reaction, calculated
  from \cite{GravenLong1954}, for three reduction states ${\rm R/O}$$=$$1.12,
  2.0, 3.0$ and an Earth/Venus-like H, C, and N reservoir. We overplot the
  cooling curves for the three R/O values during and after termination of
  planetary accretion, assuming that the accretion luminosity drops on a
  time-scale of 0.1 Myr and that H$_2$ and CO$_2$ are subsequently removed (by
  XUV irradiation and carbonate burial, respectively) on a 10 Myr time-scale.
  The freeze-out temperature of the water-gas shift reaction lies generally in
  the interval between 500 K and 550 K.}
  \label{f:wgsr}
\end{figure}

Regarding the second route to CH$_4$ formation, abiotic serpentinization by
reaction of water with the primitive mantle, kinetic models that include
formation of methane by serpentinization and destruction by photolysis find very
low CH$_4$ concentrations, at the ppm level, in oxidizing atmospheres
\citep{Guzman-Marmolejo+etal2013} and extremely low HCN concentrations near the
surface at the $10^{-16}$--$10^{-14}$ level \citep{Pearce+etal2022}. We discuss
the kinetic freeze-out temperatures of CH$_4$, NH$_3$ and HCN in more
details in Section \ref{s:fixation}.

\subsection{Chemical equilibrium of CH$_4$- and NH$_3$-free atmospheres}

\begin{figure}
  \begin{center}
    \includegraphics[width=0.9\linewidth]{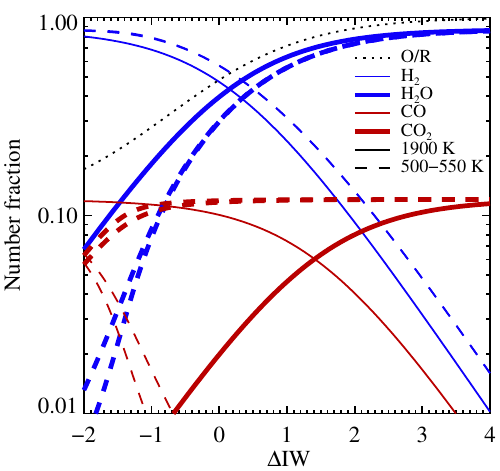}
  \end{center}
  \caption{The equilibrium mixing ratios of H$_2$, H$_2$O, CO and CO$_2$ for an
  atmosphere with 100 bar pressure and C/H=0.07 buffered in oxygen by the
  magma ocean at $T=1900\,{\rm K}$ (full lines) and after cooling down to
  $T=500\,{\rm K}$ or $T=550\,{\rm K}$ (dashed lines) under conservation of
  elements. We exclude here CH$_4$ and NH$_3$ from the equilibrium. The
  composition is given as a function of the magma oxygen fugacity relative to
  the iron-w\"ustite buffer (IW), $\Delta {\rm IW} = {\rm log}(P_{\rm O_2}/{\rm
  bar})-{\rm log}(P_{\rm IW}/{\rm bar})$. The dotted line shows ${\rm O/R} =
  ({\rm R/O})^{-1}$; the atmosphere becomes strongly oxidizing for $\Delta {\rm
  IW} \gtrsim 1\ldots2$. The water-gas shift reaction efficiently destroys CO by
  reaction with H$_2$O when the temperature is decreased. H$_2$O and CO clearly
  cannot coexist in significant quantities at low temperatures; CO$_2$ becomes
  the dominant C-carrier even at low oxygen fugacities.}
  \label{f:magma_atmosphere_cool}
\end{figure}

We start by analysing the case where CH$_4$ and NH$_3$ are assumed to be
efficiently dismantled by photolysis. The equilibrium ratios of H$_2$, H$_2$O,
CO and CO$_2$ change with the ambient temperature. Particularly, the so-called
water-gas shift reaction,
\begin{equation}
  {\rm CO} + {\rm H}_2{\rm O} \rightleftharpoons {\rm CO}_2 + {\rm H}_2 \, ,
\end{equation}
shifts to the right with decreasing temperature
\citep{Sato+etal2004,Bustamante+etal2004,Sossi+etal2020,Liggins+etal2022}.
Hence, while CO and H$_2$O can exist in equilibrium at the high temperatures
experienced directly after the crystallisation of the magma ocean, O is
transferred from H$_2$O to CO$_2$ at lower temperatures. We perform the chemical
equilibrium calculations at a temperature of 500 K; this is the approximate
surface temperature that we obtained in the calculations by
\cite{Johansen+etal2023b} for an Earth-mass planet with 100 bar of oxidizing
atmosphere (which included the effect of water condensation on the adiabatic
lapse rate); this approximate temperature is also in good agreement with
previous studies \citep{AbeMatsui1988,Kasting1988}. We demonstrate the effect of
increasing the temperature by 10\%, to 550 K, in Figure
\ref{f:magma_atmosphere_cool}. We ignore the condensation of water to form the
first oceans. At the assumed surface temperature of 500 K, the saturated vapor
pressure of water vapor is approximately 30 bar, while the maximum water
pressure based on the composition of our model is 75 bar. We checked that
condensation of the excess water vapor does not change our calculated chemical
equilibrium significantly.

In the top panel of Figure \ref{f:simple_atmosphere} we show the relative
abundances $2\,{\rm O}_2$/O, $2\,{\rm H}_2$O/H and CO$_2$/C as a function of the
reduction parameter R/O and C/H at a total pressure of 100 bar and temperature
of 500 K. As expected, O$_2$ becomes an important O carrier for ${\rm
R/O}$$<$$1$, irrespective of C/H, but vanishes for ${\rm R/O}$$\geq$$1$.
Increasing R/O further causes first H$_2$O to convert to H$_2$ while CO$_2$/C is
always at its maximal value of unity. Remarkably, the equilibrium chemistry
dictates that only when the atmosphere is already completely depleted of H$_2$O
will any further increase of R/O allow CO to co-exist with CO$_2$. This
transition happens at ${\rm C/O}$$=$$1/2$, since higher values of C/O imply an
oxygen budget insufficient to fully oxidize all C to CO$_2$. The atmosphere
finally becomes ultrareducing at ${\rm C/O}$$\geq$$1$. We do not model this
H$_2$O- and CO$_2$-free state here, since such a reducing composition is not
plausible for realistic magma ocean oxidation states \citep{Hirschmann2022}.

The equilibrium of the water-gas shift reaction precludes coexistence of H$_2$O
and CO at low temperatures. We nevertheless need to analyse the speed (or
kinetics) of the reaction, to assess the minimum temperature range at which the
equilibrium is reached during the relevant evolution time-scale of the
atmosphere. The slowing of chemical kinetics at low temperature can cause an
atmosphere to maintain the composition of an effective equilibrium temperature
that is much higher than the actual temperature. The kinetics of the water-gas
shift reaction were studied experimentally by \cite{GravenLong1954}. They found
the following production rate of CO$_2$ given known concentrations of CO, H$_2$O
and
H$_2$,
\begin{eqnarray}
  \frac{{\rm d}[{\rm CO_2}]}{{\rm d}t} &=& 5.0 \times 10^{12}\,{\rm s^{-1}}\,
  \exp\{-2.83 \times 10^5\,{\rm J\,mol^{-1}}/(R_{\rm gas} T)\} \nonumber \\
  & & \times \, [{\rm CO}]^{0.5} [{\rm H_2O}] \times (1+1.2 \times 10^4 [{\rm
  H_2}])^{-1/2} \, ,
  \label{eq:gl54}
\end{eqnarray}
where brackets [] denote concentration in moles per liter and $R_{\rm gas}$ is
the universal gas constant. We ignore from here the reduction of the reaction
rate by the presence of H$_2$; we will later show that any H$_2$ produced in the
water-gas shift reaction will escape quickly from the planet due to stellar
irradiation. The CO destruction time-scale is simply expressed as
\begin{eqnarray}
  \tau_{\rm CO} &=& -\frac{[{\rm CO}]}{{\rm d}[{\rm CO}]/{\rm d}t} 
  = 2.0 \times 10^{-13} \, {\rm s}\, \times \nonumber \\
  && \exp\{2.83 \times 10^5\,{\rm J\,mol^{-1}}/(R_{\rm gas} T)\} [{\rm
  CO}]^{0.5} [{\rm H_2O}]^{-1}\, ,
\end{eqnarray}
where we took ${\rm d}[{\rm CO}]/{\rm d}t = - {\rm d}[{\rm CO_2}]/{\rm d}t$ from
equation (\ref{eq:gl54}) by virtue of C conservation. We assume now that the
whole atmospheric column is coupled to the surface layer by rapid convection,
so that the chemical reactions affecting the entire column effectively proceed
at the surface temperature. The CO depletion time-scale, in millions of years,
for a model with a terrestrial H reservoir and venusian C and N reservoirs
chosen to represent Earth's primordial atmosphere, is shown in Figure
\ref{f:wgsr}. The time-scale rises steeply from 1 yr at 700 K, 100 yr at 600 K,
1 Myr at 550 K to 1 Gyr at 500 K.

We overplot in Figure \ref{f:wgsr} the cooling curves for atmospheres with three
different values of the oxidation state R/O (1.12, 2.0 and 3.0).  We wish to
compare these reaction rates with the rate of cooling of the atmosphere. For
computing cooling curves, we use the atmosphere equilibrium model of
\cite{Johansen+etal2023b}, which integrates the atmosphere from the surface to
the photosphere under hydrostatic equilibrium and appropriately choosing either
the radiative or the convective temperature gradient. The code uses a grey
approach to radiative transfer, with the opacity of H$_2$O and CO$_2$
proportional to pressure ($\kappa_{\rm H_2O}=1.0\,{\rm m^2\,kg^{-1}}$ at 1 bar
pressure for H$_2$O and $\kappa_{\rm CO_2}=0.01\,{\rm m^2\,kg^{-1}}$ for
CO$_2$). The latent heat of water is taken into account in the convective
temperature gradient \citep{Leconte+etal2013}. We initially decrease the
accretion luminosity (i.e., the energy release rate of the accreted material)
over a time-scale of $10^5$ yr to mimic the dissipation of the protoplanetary
disc and the decline of the pebble accretion rate in the terrestrial planet
region, until stellar irradiation becomes the dominant heat
source\footnote{Mantle cooling is also included in the model, using the discrete
interior structure model of \cite{Johansen+etal2023a}, but heat transport from
surface to atmosphere becomes insignificant after the magma ocean has
crystallized.}. We set the planetary albedo to $A=0.5$ and the stellar
luminosity to $0.7$ times the modern solar luminosity. Both H$_2$ are CO$_2$
are subsequently removed on a time-scale of 10 Myr, to mimic loss of H$_2$ by
XUV irradiation (see Section \ref{s:Hloss}) and loss of CO$_2$ by sedimentation
of carbonates in the ocean.  \cite{Sleep+etal2001} derive a CO$_2$ removal
time-scale between 10 Myr and 100 Myr (which depends, among other things, on the
availability of cations to bind C in mineral form); we chose here the lower
limit to reduce the computational time of the model.

The cooling curves in Figure \ref{f:wgsr} show that the more reducing models
obtain a higher temperature in equilibrium with stellar irradiation. This is due
to the lower amounts of water in these models, which decreases the cooling
effect of moist convection. We also tested the effect of our choice of the
opacity level of CO$_2$, by comparing nominal level of $\kappa_{\rm
CO2}=0.01\,{\rm m^2\,kg^{-1}}$ at 1 bar pressure \citep{Badescu2010} and a lower
level of $\kappa_{\rm CO2}=0.001\,{\rm m^2\,kg^{-1}}$  which we found more
consistent with the temperature of modern Venus \citep{Johansen+etal2023b}. The
opacity level of CO$_2$ nevertheless has a negligible effect on the temperature
structure, which is mainly set by the opacity of H$_2$O and the moist adiabat of
the H$_2$O/CO$_2$/N$_2$ mixture. Overall, Figure \ref{f:wgsr} demonstrates that
the water-gas shift reaction freezes out at a temperature interval between 500 K
and 550 K. This value is mainly dependent on the removal time-scale of CO$_2$ in
the two reducing cases with ${\rm R/O}$$=$$2.0$ and ${\rm R/O}$$=$$3.0$. Under
all circumstances, this temperature range is low enough that CO is effectively
converted to CO$_2$ during the cooling.

\subsection{Chemical equilibrium of atmospheres with CH$_4$ and NH$_3$}
\label{s:CH4NH3}

The kinetic time-scale of equations (\ref{eq:CH4}) and (\ref{eq:NH3}) to reach
equilibrium was reported by \cite{Zahnle+etal2020} and \cite{Liggins+etal2023}.
At an initial cooling time of 100,000 yr, \cite{Liggins+etal2023} find that the
CH$_4$ equilibrium freezes out at 700 K and NH$_3$ at 1,000 K. As we demonstrate
later in our lightning heating calculations, CH$_4$ and NH$_3$ become
significant C and N hosts at those temperatures (Figure
\ref{f:reduced_molecules}). The equilibrium time-scales at 500 K rise steeply to
$\sim$$10^9$ yr for CH$_4$ and $\sim$$10^{10}$ yr for NH$_3$. Hence any
destruction of those species by photolysis will not be replenished at the likely
surface equilibrium temperature of the early atmosphere.
\begin{figure}
  \begin{center}
    \includegraphics[width=0.9\linewidth]{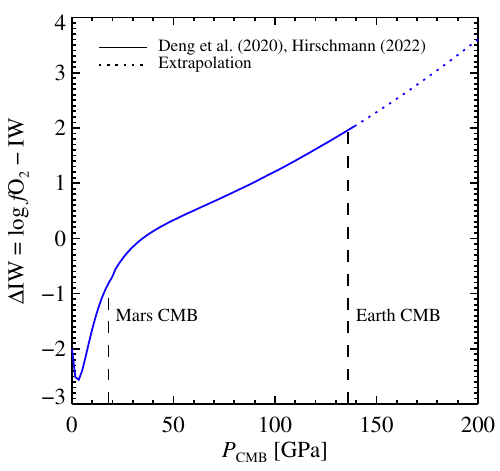}
  \end{center}
  \caption{The oxygen fugacity, relative to the iron-w\"ustite buffer (IW), at
  the surface of the magma ocean after the termination of accretion, as a
  function of the pressure at the base of the magma ocean. We make the
  calculation here following the approach of \cite{Deng+etal2020} and
  \cite{Hirschmann2022}; the calculation was done for the composition of bulk
  silicate Earth. The equation of state is valid up to 140 GPa
  \citep{Deng+etal2020} and we indicate extrapolation to higher pressures with
  dots. The relevant pressures at the core-mantle boundary (CMB) of Mars and
  Earth are shown in dashed lines.}
  \label{f:DIW_PCMB}
\end{figure}

We nevertheless analyse here a model that includes both CH$_4$ and NH$_3$ in the
chemical equilibrium in the bottom panel of Figure \ref{f:simple_atmosphere}. We
use here the FastChem package, described in Section \ref{s:yield}, to find the
chemical equilibrium. FastChem does not include water condensation, so we
ignore this effect here, as we also did for the calculations in Section 2.3.
The nature of reducing atmospheres changes character drastically compared to the
nominal model (shown in the top panel of Figure \ref{f:simple_atmosphere}), with
CH$_4$ and NH$_3$ taking over as the dominant carriers of C and N for high R/O
and low C/H.  Atmospheres with a higher C/H value show significant amounts of
CO$_2$ co-existing with CH$_4$. H$_2$O is here present at all values of R/O, in
contrast to the nominal model, since H$_2$ is oxidized by the excess O released
when C is transferred from CO$_2$ and CO to CH$_4$. We demonstrate in Section
\ref{s:Hloss} that XUV-driven mass loss leads to efficient self-oxidation in
both the nominal chemical equilibrium model without CH$_4$ and NH$_3$ and in the
model analyzed here that includes these molecules; hence the main result of our
paper is relatively insensitive to whether or not CH$_4$ and NH$_3$ are
destroyed by photolysis.

\subsection{The oxygen fugacity of the magma ocean}

We have so far considered R/O to be a free parameter in our chemical
equilibrium models, but in reality this value is set by outgassing of C, H and O
from the magma ocean. We calculate the equilibrium composition of the outgassed
atmosphere over the magma ocean at a temperature of $T=1900\,{\rm K}$, for a
range of magma ocean oxygen fugacities relative to the iron-w\"ustite buffer
reaction ${\rm Fe} + (1/2)\,{\rm O_2} \rightleftharpoons {\rm FeO}$
\citep{Ortenzi+etal2020,Hirschmann2021}. The oxygen fugacity (i.e.\ the chemical
activity or effective partial pressure of the gas, measured in bar) relative to
a strongly oxidizing magma ocean with 100\% FeO is denoted in logarithmic form
by $\Delta {\rm IW}$ \citep{Huang+etal2021}; we vary $\Delta {\rm IW}$ from -2
(strongly reducing) to +4 (strongly oxidizing). Figure
\ref{f:magma_atmosphere_cool} shows the composition of the outgassed atmosphere
as a function of $\Delta {\rm IW}$ at a total pressure of 100 bar.  The figure
also shows the composition when the temperature is lowered from the initial
$T=1900\,{\rm K}$ to either $T=500\,{\rm K}$ or $T=550\,{\rm K}$ (the likely
interval of the freeze-out temperature, see Figure \ref{f:wgsr}), under
conservation of H, C, N and O and under the assumption that CH$_4$ and
NH$_3$ are irrevocably destroyed by photolysis. The water-gas shift reaction
very efficiently destroys CO, so that CO$_2$ becomes the dominant C-carrier even
at low oxygen fugacities.

The oxygen fugacity of the magma ocean relative to the iron-w\"ustite buffer
is expressed as function of the magma composition as
\begin{equation}
  \Delta {\rm IW} = 2 \log \left( \frac{a_{\rm FeO}^{\rm
  (sil)}}{a_{\rm Fe}^{{\rm (met)}}} \right) \, .
  \label{eq:DIW}
\end{equation}
This expression is valid when FeO in the magma is in balance with free metal
\citep{Huang+etal2021}. Here $a$ denotes the chemical activity of the dissolved
species (approximately equal to the mole concentration in the relevant solvent):
$a_{\rm FeO}^{\rm (sil)}$ denotes activity of FeO in the silicate melt and
$a_{\rm Fe}^{\rm (met)}$ denotes activity of iron in the descending metal
droplets. Earth's mantle FeO fraction (8\%) implies differentiation at IW-2,
while Mars likely differentiated at IW-1.5 due to this small planet's higher FeO
fraction of 14-18\% \citep{YoshizakiMcDonough2020, Hirschmann2022,
Johansen+etal2023a}. The reaction FeO + (1/4) O$_2$ $\rightleftharpoons$
FeO$_{1.5}$ within the silicate melt nevertheless becomes increasingly important
at high pressures. As planetary accretion terminates and free metal sediments
from the magma ocean, the current leading hypothesis to explain the high
oxidation state of Earth's mantle is that the ratio of Fe$^{3+}$ (in
FeO$_{1.5}$) to Fe$^{2+}$ (in FeO) was set by the conditions at the core-mantle
boundary (CMB) where a high Fe$^{3+}$ fraction is necessary for the silicate
liquid to remain in equilibrium with core metal \citep{Armstrong+etal2019}.
Efficient convective flows, driven by the hot lower mantle that has been heated
by sinking of metal to the core, dictate a constant Fe$^{3+}$/Fe$^{2+}$ ratio
from the CMB up the magma ocean surface \citep{Armstrong+etal2019}. The increase
in the mean Fe oxidation state near the surface yields a high oxygen fugacity
there and hence the outgassing of an oxidizing atmosphere. We follow here the
calculations of \cite{Deng+etal2020} and \cite{Hirschmann2022}, who used
thermodynamical modelling of the equation state of FeO and FeO$_{1.5}$ valid up
to a pressure of approximately 140 GPa, to calculate the oxygen fugacity of the
magma ocean.
\begin{figure}
  \begin{center}
    \includegraphics[width=0.9\linewidth]{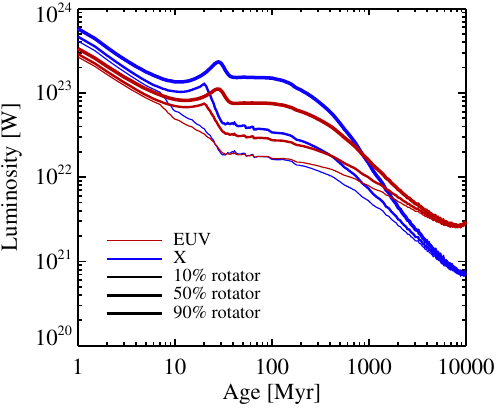}
  \end{center}
  \caption{The evolution of the X-ray luminosity and the EUV luminosity
  with age for a solar-mass star, from the tables provided by
  \cite{Johnstone+etal2021}. We show results for stars with positions of 10\%,
  50\%, 90\% along the normalized initial distribution of rotation speeds.  The
  X-ray luminosity drops by a factor of approximately 400-600 between 1 and
  5,000 Myr of stellar age, while the EUV luminosity drops by a factor of
  approximately 100.}
  \label{f:XUV_luminosity_time}
\end{figure}

The calculated oxygen fugacity at the surface is shown as a function of pressure
in Figure \ref{f:DIW_PCMB}, for an Earth-like core-mantle boundary oxygen
fugacity at IW$-2$. We base the calculations here on the composition of
Earth, since the mantle equation of state for the Mars composition in
\cite{Hirschmann2022} is fit to pressures only up to the Mars core-mantle
boundary; we checked that up to this pressure, the results based on either the
Earth or the Mars composition yield relatively similar surface oxygen
fugacities. Figure \ref{f:DIW_PCMB} shows that the magma ocean of a Mars-mass
planet has a surface oxygen fugacity of IW-1 after the termination of accretion
(which would rise to IW-0.5 for the actually higher FeO fraction of Mars), while
an Earth-mass planet outgasses its atmosphere under very oxidizing conditions at
IW+2 \citep{Sossi+etal2020}. These conditions are slightly less oxidizing than
the results of \cite{Armstrong+etal2019} who used an equation of state valid at
low pressures \cite[see discussion in][]{Hirschmann2022}.
\begin{figure*}
  \begin{center}
    \includegraphics[width=0.8\linewidth]{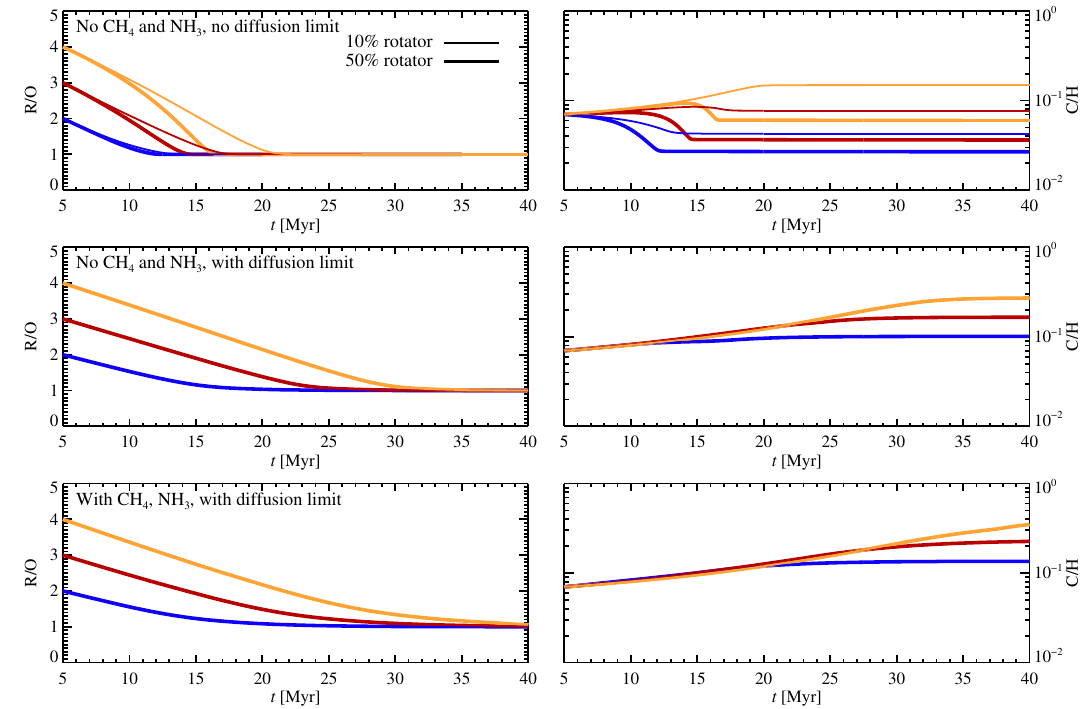}
  \end{center}
  \caption{The evolution of R/O (left panels) and C/H (right panels) due to
  mass loss by XUV irradiation, for models without CH$_4$ and NH$_3$ (top and
  middle panels) and a model including those molecules (bottom panel). In the
  top panel we do not apply a diffusion limit to the H escape flux. The
  atmospheres undergo self-oxidation (moving towards R/O=1) within 10--20 Myr
  for the first case and within 20--40 Myr for the two cases including the
  diffusion limit, due the escape of H. This H flux drags out the heavier C, N
  and O atoms as long as the flux is sufficiently high. The higher H flux in the
  top panel leads to decrease of C/H for the 50\% rotator star; this is a result
  of efficient loss of C (hosted in CO and CO$_2$) combined with inefficient
  loss of H$_2$O that is cold-trapped in the lower atmosphere. In contrast, C/H
  always increases with time in the middle and bottom panels where the diffusion
  limit is applied and heavier atoms and molecules fractionate substantially in
  the weaker escape flow.}
  \label{f:atmosphere_loss}
\end{figure*}

Extrapolating these results to rocky exoplanets with generalized masses ranging
from below Mars mass to super-Earth masses, the results in Figure
\ref{f:DIW_PCMB} predict that smaller planets should have low mantle oxygen
fugacities that correspond well to their FeO (Fe$^{2+}$) fraction, see equation
(\ref{eq:DIW}), with no increase of the FeO$_{1.5}$ (Fe$^{3+}$) fraction at the
core-mantle boundary. Planets as massive or more massive than Earth, on the
other hand, are predicted to have strongly oxidizing mantles. This extrapolation
nevertheless depends on the FeO fraction of the mantle material, since the
oxygen fugacity in Figure \ref{f:DIW_PCMB} is anchored at the value of equation
(\ref{eq:DIW}) at the core-mantle boundary. In \cite{Johansen+etal2023a}, we
demonstrated that the FeO fraction in the pebble accretion model for rocky
planet formation will be a decreasing function of the mass. This is due to the
accretion of pebbles that contain metallic iron with an intrinsically low
oxidation degree. If, on the other hand, rocky planets form by accretion of
planetesimals that experienced extensive iron oxidation by aqueous alteration
\citep{Zolensky+etal1989,Rosenberg+etal2001,Zolensky+etal2008}, then the FeO
fraction of a rocky planet could be significantly higher. Also, reaction between
FeO and H$_2$ sourced from the gas envelope accreted from the protoplanetary
disc could change the Fe-FeO-H$_2$-H$_2$O distribution significantly on both
planetary embryos and fully formed planets with magma oceans
\citep{IkomaGenda2006,Kite+etal2020,Young+etal2023}. However, even if rocky
exoplanets have a factor three lower or higher FeO fraction in their mantle
compared to Earth, then this would yield at most a change of logarithmic
core-mantle boundary oxidation state by $1$ (equation \ref{eq:DIW}), which would
reflect in a surface oxygen fugacity displaced by at most $\pm 1$ in Figure
\ref{f:DIW_PCMB}.  Changing the oxygen fugacity of an Earth-mass planet from
$\Delta {\rm IW}=2$ to $\Delta {\rm IW}=1$, by lowering its FeO mantle fraction,
would only change R/O slightly from 1.1 to 1.3 (Figure \ref{f:DIW_PCMB}); these
are still oxidizing conditions.

\subsection{Atmosphere loss and self-oxidation}
\label{s:Hloss}

Here we evolve the number of H, C, N and O atoms in the atmospheres through XUV
mass loss. The atmosphere model is assumed to instantaneously shift its chemical
equilibrium to the prevalent atomic composition during the mass loss.  We
describe the XUV mass loss rates model in detail in Appendix A. Importantly, we
assume that the mixing ratio of H$_2$O above the photosphere is equal to its
value at the photosphere (where the atmosphere temperature equals the skin
temperature), due to cold-trapping by cloud formation. We also apply a limit to
the H escape either by scaling the mass-loss efficiency with the mixing ratio of
H in the thermosphere or by imposing that the H flux can not be higher than the
diffusion rate from the lower atmosphere. For simplicity, we fix the temperature
at 500 K during the mass loss. We include now in the models without CH$_4$
and NH$_3$ the condensation of H$_2$O vapor to form oceans. Ocean condensation
is nevertheless still not included in the FastChem calculations that incorporate
those and other additional species. Since water vapor anyway experiences strong
cold-trapping in our models, the condensation of oceans does not significantly
impact atmospheric mass loss rates.

We input the evolution of the X-ray and EUV luminosity of young stars of
different initial rotation speeds from precalculated tables provided by
\cite{Johnstone+etal2021}. The initial rotation speed is, in turn, quantified by
the position of the star in the relative distribution function of rotation
speeds. The temporal decline of the X-ray luminosity and the EUV luminosity of a
solar-mass star is shown in Figure \ref{f:XUV_luminosity_time} for initial
rotation rates of 10\%, 50\% and 90\% along the normalized cumulative
distribution. Rocky planets orbiting in the habitable zone are exposed to a very
high XUV flux during the first billion years of stellar rotation slow-down, with
substantial differences between the luminosity curves in the 10--1,000 Myr
epoch.

We expose atmospheres with three initial values of R/O (=4.0, 3.0, 2.0), and
Earth/Venus-like C/H and N/H values, to XUV irradiation for up to 40 Myr after
the formation of the star. We assume that the protoplanetary disc has been
accreted and photoevaporated to oblivion after 5 Myr and that the atmosphere
thereafter quickly cools down to an equilibrium temperature of 500 K, which we
assume for simplicity to remain constant even when the atmospheric pressure and
composition change with time. We evolve the number of H, C, N and O atoms as a
function of time by integrating equations (\ref{eq:Ndoti}) and (\ref{eq:Ndots})
for the 10\% and 50\% rotator, noting that the 90\% rotator gives even more
efficient mass loss. Figure \ref{f:atmosphere_loss} shows the evolution of R/O
and C/H with time, both for the model without CH$_4$ and NH$_3$ (top and middle
panels) and for the model including those molecules (bottom panel). The
composition quickly drops towards R/O=1 for both models. This evolution takes
only 10--20 Myr for the top panel where we do not include a diffusion limit to
the H escape flux. In the middle and bottom panels, where the diffusion limit is
applied to the model without and with CH$_4$ and NH$_3$, the self-oxidation
time-scale is 20--30 Myr and 25--40 Myr, respectively, depending on the initial
value of R/O. Including the diffusion limit, mass loss rates are independent of
the initial rotation speed, because diffusion from the lower atmosphere sets the
maximal mass loss rate in that case.

The C/H ratio displayed on the right panels of Figure \ref{f:atmosphere_loss}
drops only in the top panel for the 50\% rotator without the diffusion limit,
due to the high H fluxes that can in that case efficiently drag out C (plus N
and O) with little fractionation. This leads to a decreased C/H ratio because
the bulk part of the H$_2$O component is cold-trapped and does not participate
in the atmosphere escape. In the models with the diffusion limit, C/H increases
by a factor 2--5 due to extensive fractionating in the mass loss flux. Some
representative self-oxidation tracks based on Figure \ref{f:atmosphere_loss} are
also indicated in Figure \ref{f:simple_atmosphere}.

\begin{figure}
  \begin{center}
    \includegraphics[width=0.9\linewidth]{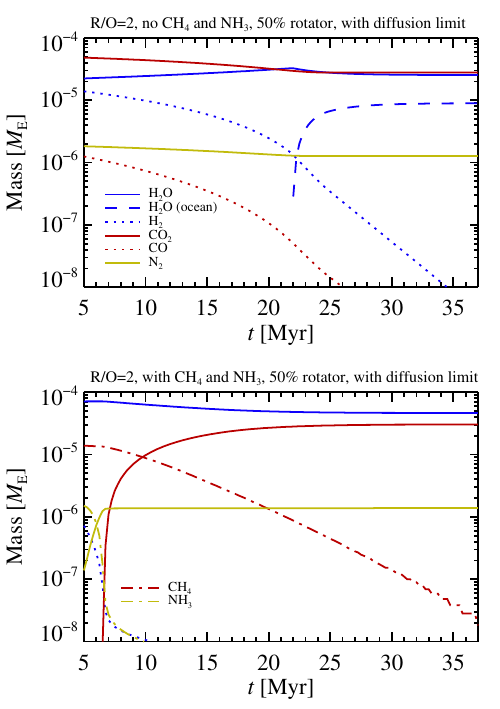}
  \end{center}
  \caption{The evolution of the molecular species in the middle model and
  bottom models of Figure \ref{f:atmosphere_loss} starting at R/O=2. In the top
  panel (no CH$_4$ and NH$_3$), the XUV irradiation leads to escape of the
  reducing H$_2$ and CO molecules within approximately 30 Myr. CO$_2$ and N$_2$
  experience some mass loss too as these molecules are dragged out together with
  the escaping H. The first oceans condense out after 22 Myr when the pressure
  of the atmosphere has dropped enough that water vapor can no longer be
  sustained at its equilibrium pressure. In the bottom panel (where we use
  FastChem to include a wider range of molecules), the initial small amount of
  H$_2$ escapes quickly and subsequent oxidation of the atmosphere is driven by
  loss of H donated by CH$_4$ destruction. The surplus C is oxidized to CO$_2$
  by reaction with water vapor, while the amount of CO formed in equilibrium is
  insignificant (at the 550--600 K surface temperature range of these reducing
  conditions, see Figure \ref{f:wgsr}).}
  \label{f:atmosphere_composition}
\end{figure}

In Figure \ref{f:atmosphere_composition} we show the evolution of the
atmospheric species in the 50\% rotator model that starts at R/O=2 and includes
the diffusion limit, for models with and without CH$_4$ and NH$_3$. For the
model without these molecules (top panel), the dominant reducing molecule is
H$_2$, with CO playing only a minor role due to the water-gas shift
reaction. Both H$_2$ and CO fall to trace levels in $\sim$$30$ Myr of mass loss
evolution. The first oceans condense out after 22 Myr when the atmospheric
pressure has fallen to a low enough value that water vapor can no longer be
present at its equilibrium pressure for the nominal mixing ratio. CO$_2$ and
N$_2$ masses fall by nearly a factor of two, as these species are dragged along
with the H outflow. The complete loss of H$_2$ nevertheless leads to an
increase in the bulk C/H ratio (see middle right panel of Figure
\ref{f:atmosphere_loss}). The model including CH$_4$ and NH$_3$ (calculated
using FastChem) loses its free H$_2$ budget very rapidly. The subsequent
oxidation of the atmosphere happens by loss of H donated by CH$_4$ destruction
in the thermosphere. The excess C is oxidized in equilibrium to CO$_2$ by
reaction with water vapor.

Overall, the mass loss calculations presented in Figure \ref{f:atmosphere_loss}
and Figure \ref{f:atmosphere_loss} show that atmospheres undergo a rapid
self-oxidation process within a few 10 Myr. This conclusions holds relatively
irrespectively of the chemical equilibrium model, variations to the mass loss
calculations and variations in the initial oxidation state of the outgassed
atmosphere. We further test the robustness of mass loss in Figure
\ref{f:atmosphere_loss_variation} of Appendix A, where we vary the efficiency of
the energy-limited mass loss and perform an additional calculation where
molecules such as CO$_2$ are assumed indestructible and hence harder to lift by
the H flux.

\subsection{Outlook on self-oxidation calculations}

The atmospheric self-oxidation calculations presented here employ an
end-member approach to modelling thermospheric escape, by assuming either (a)
complete atomization of all molecules in the thermosphere (in this section) or
(b) destruction of only H$_2$ (see Appendix A where we explore the latter
approach to the model without CH$_4$ and NH$_3$). The goals of our paper bear
some similarities to the work of \cite{Zahnle+etal2020} who considered the
evolution of impact-generated atmospheres using chemical equilibrium combined
with quenching temperatures and a kinetic photolysis model. A direct comparison
to that work is nevertheless difficult due to the very different starting
points: we begin with an atmosphere outgassed from the magma ocean while
\cite{Zahnle+etal2020} consider the reducing effect of an iron-rich impactor on
an existing atmosphere. The short cooling time-scale following an impact, of
order $\sim$$10^3$ yr, importantly motivated \cite{Zahnle+etal2020} to consider
a freeze-out (quench) temperature of 800 K for the full
``H$_2$--H$_2$O--CH$_4$--CO--CO$_2$'' system, while we demonstrate in Figure
\ref{f:wgsr} that the reducing models have equilibrium surface temperatures in
the range 550--600 K where the water-gas shift reaction time-scale is much
shorter than the cooling time of the atmosphere. At these temperatures, we
expect CO to be rapidly converted to CO$_2$ by reaction with H$_2$O. The
reduction power of the atmosphere is thus transferred in our model almost
entirely to H$_2$, which escapes most easily, while the post-impact models of
\cite{Zahnle+etal2020} exhibit significant amounts of remnant CO after the H$_2$
component of the atmosphere has escaped.

We demonstrated in this section the efficiency of the self-oxidation process for
an Earth-mass planet, but we note that rocky planets of Earth-mass and higher
will undergo self-oxidation already at the magma ocean stage, as shown in Figure
\ref{f:DIW_PCMB}, and hence outgas strongly oxidizing atmospheres. Smaller
planets undergo less magma ocean oxidation, but have higher mass loss
efficiencies and lower gravities \citep{Salz+etal2016}. A Mars-mass planet in an
Earth-like orbit will therefore undergo atmospheric self-oxidation even more
rapidly than our nominal Earth-mass planet. We show in the next section that
self-oxidation has major implications for origin of life scenarios.

\section{Fixation of nitrogen by lightning}
\label{s:fixation}

In this section, we explore lightning as a pathway to fixing both nitrogen and
carbon. By `fixation', we refer to any chemical process that transfers C or N
away from the strongly bound CO$_2$ and N$_2$ molecules into more reactive
forms. Kinetic barriers are of great importance in allowing disequilibrium
levels of molecules containing fixed N and C to persist in the atmosphere at low
temperatures. At elevated temperatures, the equilibrium host of N shifts to
include significant amounts of NO/NO$_2$ in an oxidizing atmosphere and to
HCN/HNC in a reducing atmosphere. We will discuss the relevance of these fixed
nitrogen host molecules for prebiotic chemistry in the following section.

\subsection{Molecular freeze-out temperatures}

Shifting the atmosphere locally to very high temperatures by lightning, the
equilibrium speciation of molecules (and atoms) will change. The ``memory'' of
this high-temperature equilibrium depends in turn on the approximate freeze-out
temperature, below which the reaction rates back to the low-temperature
equilibrium proceed too slowly to matter on Gyr time-scales. We discuss here the
freeze-out temperatures of NO/NO$_2$, HCN/HNC, CH$_4$ and NH$_3$.

{\it NO/NO$_2$}. The NO abundance peaks at a temperature of around 4,000 K and
freezes out between 2,000 K and 2,250 K \citep{MancinelliMcKay1988}. In Earth's
strongly oxidizing atmosphere, O$_2$ is the main donor of oxygen to create NO
from the reaction between N and O \citep{Hill+etal1980}. In atmospheres devoid
of O$_2$, O comes mainly from H$_2$O and CO$_2$
\citep{YungMcElroy1979,ChameidesWalker1981,KastingWalker1981}, forming NO
through the reactions O + N$_2$ $\rightarrow$ NO + N and N + CO$_2$
$\rightarrow$ NO + CO \citep{Navarro-Gonzalez+etal2001} and NO$_2$ by subsequent
oxidation of NO \citep{KastingWalker1981}. Lightning in Earth's modern
atmosphere produces approximately $10^9$ kg fixed nitrogen per year
\citep{Navarro-Gonzalez+etal2001}. This gives an atmospheric depletion
time-scale of approximately 6 Gyr; the main return of fixed nitrogen back to
N$_2$ is through bacterial denitrification that harnesses energy from lowering
the energy level of nitrogen back to strongly-bound N$_2$
\citep{MancinelliMcKay1988}.
\begin{figure}
  \begin{center}
    \includegraphics[width=0.9\linewidth]{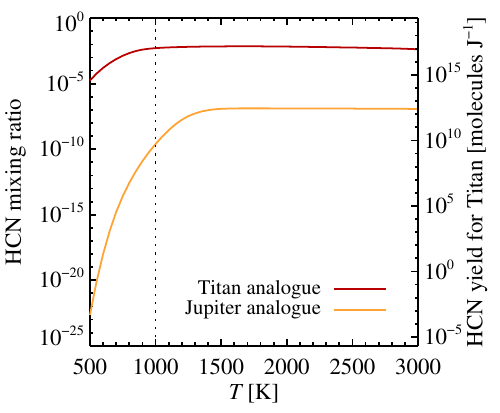}
  \end{center}
  \caption{Test of HCN concentration obtained with FastChem for a Jupiter
  analogue, where H and He abundances are reduced by a factor three in abundance
  relatively to the solar composition, and a Titan analogue consisting of 95\%
  N$_2$ and 5\% CH$_4$. The Jupiter analogue atmosphere gives a far lower HCN
  concentration than Titan; this is mainly due to the destruction of HCN by
  reaction with H$_2$ in this nearly-solar composition environment. The
  molecular yields of HCN in the Titan analogue are indicated on the right axis
  and agree well with \cite{Borucki+etal1988}.}
  \label{f:jupiter_titan_analogue}
\end{figure}
\begin{figure*}
  \begin{center}
    \includegraphics[width=0.8\linewidth]{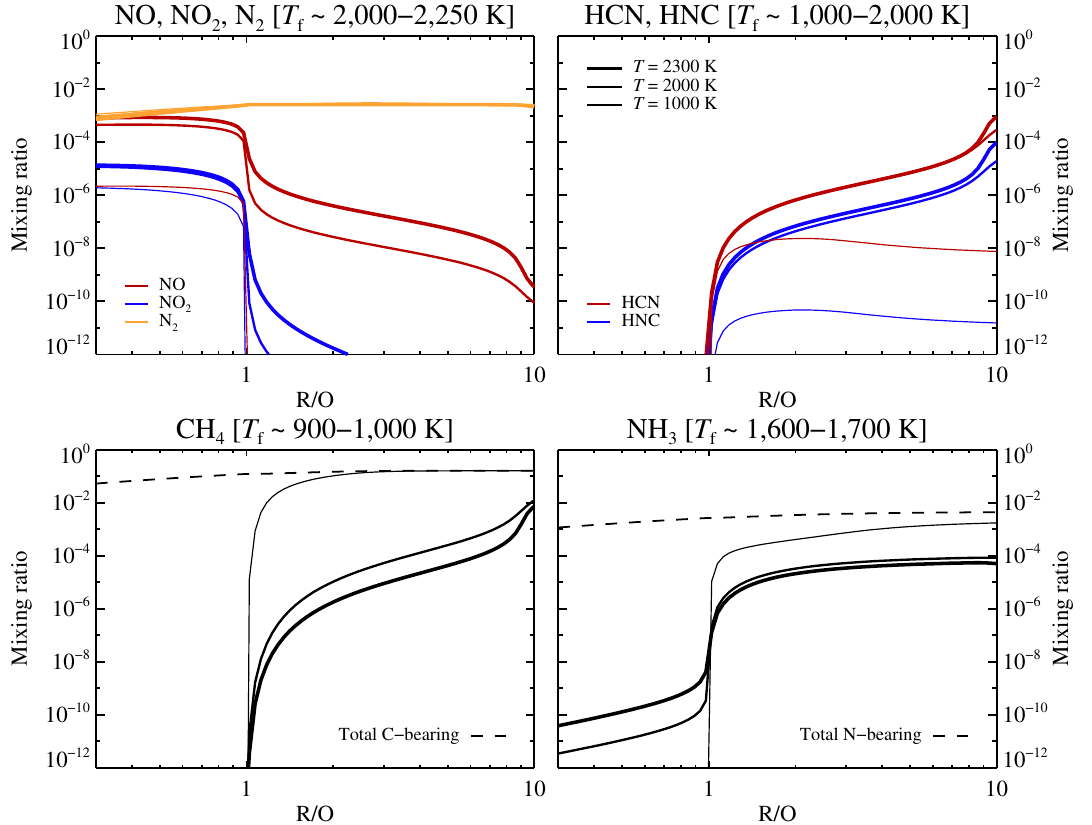}
  \end{center}
  \caption{Mixing ratio of the C and N host molecules NO, NO$_2$ and N$_2$ (top
  left); HCN and HNC (top right); CH$_4$ (bottom left); and NH$_3$ (bottom
  right) for C/H=0.07 (the approximate value for bulk silicate Earth), as
  function of R/O and at three different temperatures. The approximate
  freeze-out temperatures of the chemical kinetics are indicated in the plot
  titles. The production of NO and NO$_2$ molecules drops slowly when
  transitioning from oxidizing atmospheres with R/O$\leq$$1$ to reducing
  atmospheres with R/O$>$$1$.  With increasing R/O above unity, HCN and HNC
  concentrations in turn rise slowly. CH$_4$ becomes the main carrier of carbon
  at its freeze-out temperature for R/O$>$$1$, while NH$_3$ dominates over HCN
  and HNC as the main carrier of fixed nitrogen for all R/O$>$1.}
  \label{f:reduced_molecules}
\end{figure*}

{\it HCN/HNC}. In reducing atmospheres, HCN is produced in large quantities by
lightning heating. The lightning energy destroys N$_2$ and CH$_4$ molecules to
form the radicals N and CH$_3$ that readily combine to HCN
\citep{Pearce+etal2022}. \cite{ChameidesWalker1981} report an HCN
freeze-out temperature of 2,000--2,500 K. This contrasts with Jupiter models
that find that the HCN level freezes out at a temperature below approximately
870 K \citep{Moses+etal2010}. \cite{Borucki+etal1988} reported an HCN freeze-out
of approximately 950 K from lightning experiments and found much higher HCN
yields for their Titan atmosphere models than for Jupiter, due to destruction of
HCN by H$_2$ under Jupiter conditions \citep{Moses+etal2010}. The contrast
between the high freeze-out temperature of \cite{ChameidesWalker1981} and the
lower value favored by \cite{Borucki+etal1988} may be due to different
assumptions about the strength of lightning; indeed \cite{ChameidesWalker1981}
also report an elevated freeze-out temperature of NO at 3,000--3,500 K. We
therefore take here the freeze-out temperature of HCN to lie in the interval
1,000--2,000 K.

{\it CH$_4$}. Methane forms by consecutive reactions of CO and its products with
H$_2$ \citep{ZahnleMarley2014}. Methane is the energetically favorable C-host at
low temperatures in reducing atmospheres. The reaction ${\rm CH_3OH} + {\rm H_2}
\rightleftharpoons {\rm CH_4} + {\rm H_2O}$ nevertheless appears as bottleneck
for CH$_4$ production; this reaction becomes kinetically inhibited below a
freeze-out temperature in the range of 900-1,000 K
\citep{Moses+etal2011,ZahnleMarley2014,Sossi+etal2020}.

{\it NH$_3$}. Ammonia forms by consecutive reactions of N$_2$ and its products
with H$_2$. NH$_3$ is the dominant carrier of N in reducing conditions. The
equilibrium between NH$_3$ and N$_2$ freezes out at a temperature of 1,600-1,700
K where the NH$_3$ ratio is relatively low \citep{Moses+etal2011}.

\subsection{Heating in lightning}
\label{s:heating}

Lightning is the result of charge separation in water clouds
\citep{Christian+etal2003,Yair2012}. The low conductivity of planetary
atmospheres allows the electric field to build up to its breakdown strength,
where the electrons accelerated by the field are fast enough to cause an
ionization cascade. The rapid release of electrons and increase in conductivity
lead to charge neutralization by lightning discharge in narrow discharge
channels. The lightning channel reaches a temperature up to between 20,000 K and
30,000 K within a few microseconds. The channel subsequently cools by mixing of
cold air with the surroundings \citep{Hill+etal1980}. At high temperatures, the
chemical reactions are extremely rapid and hence chemical equilibrium is
maintained at the instantaneous temperature. As demonstrated above, many
chemical reactions nevertheless have freeze-out temperatures that are
significantly above the ambient temperature of the passively irradiated
atmosphere.

\subsection{Lightning yield tests}
\label{s:yield}

The yield (or production rate) of a given molecule $i$ (in our case, NO/NO$_2$,
HCN/HNC, CH$_4$ and NH$_3$) can be quantified as the number of molecules $\Delta
N_i$ formed per input energy $\Delta E$ \citep{MancinelliMcKay1988}. With a
molecular freeze-out temperature of $T_{\rm f}$, the yield in an atmosphere of
equilibrium temperature $T_0$ and mass $M_{\rm atm}$ is defined as
\begin{equation}
  Y_i = \frac{\Delta N_i}{\Delta E} = \frac{X_i/\mu}{H(T_{\rm f})-H(T_0)}
  \label{eq:yield}
\end{equation}
where $H = c_{\rm p} T$ is the enthalpy of the atmospheric gas (with $c_{\rm p}$
denoting the specific heat at constant pressure), $X_i$ is the molar mixing
ratio of molecule $i$ at its freeze-out temperature and $\mu$ is the mean
molecular weight of the atmospheric gas at the freeze-out temperature.

We use the FastChem code to calculate the equilibrium speciation of molecules in
the gas phase as a function of temperature, pressure and elemental abundances
\citep{Stock+etal2018,Stock+etal2022}. We test our numerical calculation of
chemical equilibrium by measuring the HCN yields produced by FastChem for
Jupiter and Titan analogues for a range of temperatures. For Jupiter, we
approximate the composition of the atmosphere as solar but with the abundances
of H and He reduced by a factor of three \citep{Guillot2005}. We consider a
pressure level of 5 bar to represent the depth of Jupiter's water cloud layer
\citep{Yair2012}. For Titan, we take an atmosphere consisting of 95\% N$_2$ and
5\% CH$_4$ and a pressure of 1.5 bar \citep{Horst2017}. We show the HCN
concentration as a function of temperature in Figure
\ref{f:jupiter_titan_analogue}. As expected, the Jupiter model produces a far
lower HCN concentration than Titan. This is partially due to the low abundance
of N relative to H in this atmosphere of nearly solar composition, but more
importantly we see the effect of HCN destruction by reaction with ambient H$_2$
\citep{Moses+etal2010}. The right axis of Figure \ref{f:jupiter_titan_analogue}
shows the yield of HCN in the Titan atmosphere. A yield of slightly above
$10^{17}$ molecules per joule of input energy agrees well with the calculation
presented for Titan presented in \cite{Borucki+etal1988}.

\subsection{Lightning on rocky planets}

We now turn to modelling lightning in the outgassed atmospheres of young rocky
planets. We fix the pressure at $P=100\,{\rm bar}$ and test temperatures of
1,000 K, 2,000 K and 2,300 K. We run simulations for a range of R/O values, but
with C/H fixed at 0.07 and N/H fixed at 0.008 (see discussion in Section
\ref{s:equilibrium}). In Figure \ref{f:reduced_molecules} we show the resulting
concentrations of NO and NO$_2$, HCN and HNC, CH$_4$ and NH$_3$ as a function of
R/O and for three different values of the temperature. The value of R/O is
clearly a key determinant for the yield of fixed N and C carriers at elevated
temperatures.  For strongly oxidizing atmospheres with R/O$\leq$$1$, mainly
NO/NO$_2$ is formed, accompanied by a tiny fraction of NH$_3$ (at the $10^{-12}$
level). For R/O$>$$1$, NO/NO$_2$ levels drop slowly, while the levels of
HCN/HNC, CH$_4$ and NH$_3$ increase.  CH$_4$ becomes the dominant carrier of
carbon at its freeze-out temperature of 1,000 K, while NH$_3$ dominates as
carrier of fixed nitrogen for R/O$>$1. The relatively low concentration of HCN
found in these models (at the 1--1,000 ppm level) is due to the presence of
large amounts of H$_2$ for R/O$>$$1$, similar to what we observed for the
Jupiter model in Figure \ref{f:jupiter_titan_analogue}. We have experimented
with larger values of C/H and observed an increased HCN production at the
expense of a decrease in NH$_3$.

\section{Discussion}

\subsection{NO yields in the early atmosphere}
\label{s:NOyield}

\begin{figure}
  \begin{center}
    \includegraphics[width=0.9\linewidth]{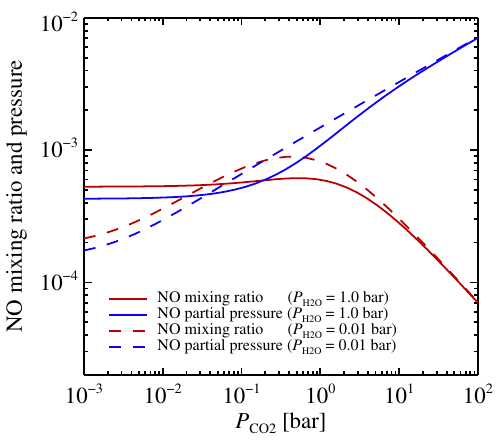}
  \end{center}
  \caption{The NO mixing ratio and partial pressure for atmosphere models with a
  range of CO$_2$ partial pressures ($x$-axis), a fixed partial pressure of
  N$_2$ (0.8 bar) and two pressure values of H$_2$O (1 bar and 0.01 bar). The
  temperature is considered here to be $T=2000$ K, representing the freeze-out
  temperature of NO. The NO mixing ratio increases significantly as the CO$_2$
  partial pressure is decreased. Below 1 bar CO$_2$ pressure, the mixing
  ratio reaches a plateau for $P_{\rm H_2O}=1\,{\rm bar}$ and even falls for
  $P_{\rm H_2O}=0.01\,{\rm bar}$, as lightning energy is then spent on heating
  H$_2$O and N$_2$ rather than CO$_2$. The partial pressure falls considerably
  over the considered interval of CO$_2$ pressure, mainly due to the decrease in
  the total oxygen budget carried by CO$_2$ for NO production.}
  \label{f:Earth_atmosphere}
\end{figure}

The mixing ratios in Figure \ref{f:reduced_molecules} were calculated at a fixed
total pressure of $P=100\,{\rm bar}$. We present now an additional set of
simulations performed with FastChem where we vary the partial pressure of
CO$_2$, while keeping the partial pressure of nitrogen fixed at $P_{\rm
N_2}=0.8\,{\rm bar}$ and the partial pressure of water vapor fixed at either
$P_{\rm H_2O}=1\,{\rm bar}$ or $P_{\rm H_2O}=0.01\,{\rm bar}$.  These represent
the equilibrium vapor pressure in atmospheres at temperature level
$100^\circ{\rm C}$ and a more temperate $20^\circ{\rm C}$, respectively. In
Figure \ref{f:Earth_atmosphere} we show the mixing ratio as well as the partial
pressure of NO at a temperature of $T=2000$ K. The mixing ratio increases
significantly with lower CO$_2$ pressure, mostly driven by the decrease in the
total number of molecules when decreasing the pressure, reaching a plateau
or a slight decline below $P_{\rm CO_2}=1\,{\rm bar}$, below which H$_2$O and
N$_2$ dominate the pressure.  The partial pressure of NO, which represents the
total NO mass, drops by 1--2 orders of magnitude over a CO$_2$ pressure drop of
four orders of magnitude, as the oxygen source for NO switches from CO$_2$ to
H$_2$O.

We use now the yield equation (\ref{eq:yield}), together with an NO
concentration from Figure \ref{f:Earth_atmosphere} of $C_{\rm NO}$$\sim$$7
\times 10^{-5}$ at 100 bar CO$_2$ pressure, to estimate an NO yield of $Y_{\rm
NO}$$\sim$$ 4 \times 10^{14} \,{\rm J^{-1}}$ (NO molecules produced per Joule)
in an early atmosphere dominated by 100 bar of CO$_2$. The yield depends only on
the mixing ratio and thus increases to $Y_{\rm NO}$$\sim$$3 \times 10^{15}
\,{\rm J^{-1}}$ for lower CO$_2$ partial pressure. This estimate agrees well
with the plateau at high CO$_2$ mixing ratios found in lightning laboratory
experiments by \cite{Navarro-Gonzalez+etal2001} who considered a CO$_2$-N$_2$
mixture at a fixed pressure level of 1 bar.

The rate of lightning strikes on modern Earth has been measured at a level of
44$\,{\rm s^{-1}}$ \citep{Christian+etal2003}. With a mean energy of
$\sim$$10^9\,{\rm J}$ per lightning strike \citep{ChybaSagan1991}, this results
in an energy dissipation rate of $4 \times 10^{10}\,{\rm W}$ or approximately
$10^{18}\,{\rm J\,yr^{-1}}$. This lightning energy is ultimately derived from
the energy carried through the atmosphere by convection
\citep{Borucki+etal1982}; hence we can assume that the lightning energy
dissipation rate on young Earth was not substantially different from the modern
value (and hence was, to first order, not affected by effects such as the
Earth's higher primordial rotation frequency). This leads to an estimated NO
production of $\sim$$10^{26}\,{\rm s^{-1}}$ under early Earth conditions. At
this rate, it would have taken $\sim$70 Gyr to process the entire N$_2$ budget
present in Earth's modern atmosphere into NO.  However, as discussed above, the
dense CO$_2$ atmosphere is not expected to have persisted for more than a few
100 Myr.  The drop of the CO$_2$ partial pressure to less than 0.1 bar, inferred
to been reached by the beginning of the Archean eon 4 Gyr ago
\citep{CatlingZahnle2020}, together with a drop in the partial pressure of
H$_2$O vapor at milder surface temperatures would have lowered the lightning
yields substantially and perhaps triggered the evolution of bacterial nitrogen
fixation \citep{Navarro-Gonzalez+etal2001}.

\subsection{Delivery of fixed nitrogen by interplanetary dust particles}

Interplanetary dust particles (IDPs) recovered on Earth exhibit typical
concentrations of $C_{\rm N}$$\sim$$10$--$100$ ppm in organic nitrogen
\citep{Marty+etal2005}. These values lie within the range of the N concentration
in ordinary chondrites and enstatite chondrites \citep{Grewal+etal2019}.
Atmospheric entry heating of organics to more than 600--800$^\circ$C
nevertheless leads to alteration and loss of organics \citep{Riebe+etal2020};
only IDPs in the mass range $10^{-15}$--$10^{-9}\,{\rm kg}$ maintain their
organics against UV photolysis (lower range) and silicate evaporation
(upper range) \citep{Anders1989,ChybaSagan1992}. Terrestrial weathering could
also have contributed significantly to nitrogen loss. Hence, the
pre-atmospheric-entry nitrogen concentration level may have been closer to 1,000
ppm \citep{Hashizume+etal2000,Marty+etal2005}, with some grains of cometary
origin reaching even higher levels of nitrogen concentration
\citep{Aleon+etal2003}.

Pyrolysis (i.e.\ thermal decomposition) of complex organic matter during
atmospheric entry transfers the nitrogen to simple compounds such as NH$_3$
\citep{Nakano+etal2003} and HCN \citep{OkumuraMimura2011}. Hence IDPs delivered
both intact organics and simple N-carrying compounds to the young Earth. Both
could have served as a source of fixed nitrogen needed by an emerging biosphere
that exploited geological serpentinization reactions as a source of H$_2$ to fix
C from atmospheric CO$_2$. NH$_3$ is nevertheless photolysed at even longer
wavelengths than CH$_4$ and transforms to N$_2$ in a CO$_2$-rich atmosphere
\citep{KuhnAtreya1979,Kasting1982} on a time-scale that is much shorter than the
CH$_4$ photolysis time-scale from \cite{Kasting2014} repeated here in equation
(\ref{eq:tCH4}).

The current flux of IDPs at the top of the atmosphere is estimated at $F_{\rm
IDP} = 1.5-4.5 \times 10^7\,{\rm kg\,yr^{-1}}$
\citep{Peucker-EhrenbrinkRavizza2000}. That gives a nitrogen delivery rate from
IDPs of
\begin{equation}
  r_{\rm N,IDP} =  4 \times 10^{22}\,{\rm s^{-1}}\,\left(
  \frac{F_{\rm IDP}}{3 \times 10^7\,{\rm kg\,yr^{-1}}} \right) \left(
  \frac{C_{\rm N}}{1,000\,{\rm ppm}} \right) \, .
  \label{eq:rNIDP}
\end{equation}
This is approximately three orders of magnitude lower than the internal nitrogen
fixation on early Earth calculated in Section \ref{s:NOyield}.
\cite{ChybaSagan1992} discuss evidence that the IDP delivery rate has been
relatively constant in time back to at least 3.6 Gyr ago, based on analysis of
IDPs from lunar soils. There is currently no known direct way to infer the IDP
flux further back in time, but \cite{ChybaSagan1992} nevertheless estimated that
the IDP flux could have been 3-4 orders of magnitude higher at the earliest
epochs of the history of Earth. We therefore suggest that delivery of
fixed nitrogen by IDPs was likely comparable in rate to fixation by lightning on
Hadean Earth.

In Appendix B we calculate the delivery rate of fixed nitrogen from asteroid
bombardment. As with interplanetary dust particles, the asteroid impact rate
during the early Hadean is very uncertain, but it was likely orders of magnitude
higher than today. Asteroid impacts are nevertheless prone to conversion of the
fixed N present in their organics into N$_2$ in the immense impact heat (as
demonstrated in Figure \ref{f:reduced_molecules}), particularly if the impactor
material mixes with the oxidizing atmosphere during the cooling.

\subsection{Relevance of our results for origin-of-life scenarios}

Life-as-we-know-it requires plenty of fixed nitrogen (often incorporated in
biomolecules as an -NH$_3$ amino group) since N is an important constituent part
of central biomolecules such as proteins, nucleic acids, and many cofactors.
Because of this, all different proposed pathways to life’s emergence include
nitrogen as one of the focal points. For instance, there are two main known
prebiotic pathways for the synthesis of amino acids. The first one relies on the
chemistry of nitriles, and uses HCN in a reaction mechanism known as Strecker
synthesis where amino acids are produced \citep{Ruiz-Bermejo+etal2013}.  This
mechanism -- which is chemically unrelated to that used by extant biology -- is
thought to be responsible for at least a portion of amino acids found in
extraterrestrial environments such as asteroids and comets
\citep{Giese+etal2022}.  The alternative prebiotic mechanism, which is
chemically analogous to how cells synthesize amino acids, occurs through
reductive amination or transamination of simple ketoacids
\citep{Mayer+etal2021,Harrison+etal2023}. The prebiotic synthesis of nucleotides
has been historically dominated by HCN-related approaches
\citep[e.g.][]{Powner+etal2009,Teichert+etal2019}. However, there has recently
been advances on the prebiotic synthesis of nucleobases using
biologically-relevant simple precursors such as carbamoyl phosphate
\citep{Yi+etal2022}, which are HCN-independent and thus fully compatible with a
CO$_2$-dominated atmosphere.

Our results are relevant for the state-of-the-art in origin of life studies
mainly by pointing out the most abundant forms of N and C that were available
for prebiotic chemistry: oxidized NOx species dissolved in the oceans (producing
NH$_3$ after interaction with marine hydrothermal environments locally rich in
H$_2$) as well as CO$_2$. In contrast, the synthesis of HCN requires an
atmosphere that is rich in already-fixed carbon (e.g., CH$_4$) or free hydrogen
(H$_2$), none of which are abundant according to our model. Hence, we
propose that the prebiotic chemistry of N that was globally available on the
Hadean Earth would have relied on reductive amination of small organics
\citep[themselves probably derived from CO$_2$ reduction by H$_2$,
see][]{Hudson+etal2020,Beyazay+etal2023} by NH$_3$.  These geochemical
mechanisms, which echo what we see in modern biology, aid in closing the gap
between early geochemistry and ancient biochemistry.

The dichotomy between a submarine versus surficial origin of life has been a
main topic of discussion during the past few decades, with HCN-dominated
chemistry usually invoked for surficial environments, and CO$_2$/NOx-dominated
chemistry for submarine ones. This environmental division nevertheless does
not necessarily correspond exactly to the two distinct chemical pathways, both
of which are technically possible in each scenario \citep[e.g.\ reactivity of
HCN has also been explored at alkaline hydrothermal vents,
see][]{Villafane-Barajas+etal2021}. Hence, our results showing a primitive
CO$_2$-dominated atmosphere do not specifically preclude either geographic
locale.  Instead, the early dominance and availability of CO$_2$ demonstrated
here is suggestive of a prebiotic chemistry based on carbon and nitrogen
fixation in a way reminiscent of ancient biology (CO$_2$- and NH$_3$-based). It
is worth noting that HCN-based prebiotic chemistry would still be plausible in
the Hadean Earth after large impactors \citep{Itcovitz+etal2022}, however
whether these transient reducing conditions prevailed for long enough for a
putative HCN-based life to emerge remains an open question.

\subsection{Implications for exoplanets}

Atmospheres of rocky planets become oxidizing both from magma ocean processes
that increase the Fe$^{3+}$ fraction in the mantle
\citep{Frost+etal2004,Armstrong+etal2019} and from XUV mass loss that
efficiently removes any reducing H that is not bound in cold-trapped H$_2$O
(this work). From that perspective, rocky planet atmospheres should always
obtain an oxidation state of R/O$\approx$1. Rocky planets close to their host
star may subsequently undergo additional destruction and loss of H$_2$O during a
later run-away greenhouse process \citep{Way+etal2016,Turbet+etal2021},
potentially leading to abiotic pile-up of O$_2$ \citep{LugerBarnes2015} and thus
Earth-like oxidation conditions with R/O$<$$1$.

We worked here with Earth/Venus-like C/H and N/H ratios in the atmospheres
and an Earth-like mantle FeO fraction (8\%), but we expect the results to be
relatively independent of these parameters unless the chemistry becomes
C-dominated with C/O$>$1.  Planets more massive than Earth can accrete a primary
atmosphere of H$_2$ and He on top of their outgassed atmosphere
\citep{IkomaHori2012}. We did not consider any interactions between the
H$_2$/H$_2$O in such an envelope and the Fe/FeO contents of the magma ocean
\citep{KimuraIkoma2020,KiteSchaefer2021,SchlichtingYoung2022,Young+etal2023}.  A
realistic span of FeO mantle fractions from three times below Earth's value to
three times above will change the oxygen fugacity by one logarithmic unit up or
down, but this would not change the picture that low-mass planets outgas
reducing atmospheres and high-mass planets outgas oxidizing atmospheres, nor
would it change our conclusion that reducing atmospheres undergo rapid
self-oxidation by XUV-driven mass loss. We also fixed the initial atmosphere
pressure at 100 bar. Higher atmosphere masses would be much harder to erode by
XUV irradiation. Mini-Neptunes that maintain either a primordial gas envelope or
an equally massive outgassed atmosphere, nevertheless, are unlikely to be
habitable, to an Earth-like biosphere at least, since the blanketing effect of
the envelope keeps the magma ocean from crystallizing
\citep{Benneke+etal2019,Tsiaras+etal2019,KiteBarnett2020,Kreidberg+etal2022}.

We note that a possible way to override the primordial self-oxidation processes
would be through late delivery of significant amounts of reduced material
by asteroid impacts \citep{Zahnle+etal2020,Pearce+etal2022,Wogan+etal2023} after
a few hundred million years, when the XUV luminosity of the young star has faded
enough to slow down the H$_2$ loss. The likelihood that such extensive
bombardments occur in a planetary system is nevertheless hard to assess, since
it depends on the dynamics of one or more giant planets \citep{MartinLivio2021}
as well as on the masses and locations of remnant planetesimals rich in
unoxidized, metallic iron.

\section{Conclusions}

In this paper we have studied molecular speciation in the atmospheres of rocky
planets as they cool down after the magma ocean phase. We have particularly
focused on the molecular host of nitrogen, since this has strong implications
for origin of life scenarios. Our main findings can be summarized as follows:
\begin{enumerate}
  \item The water-gas shift reaction in combination with loss of H from H$_2$,
  CH$_4$ and NH$_3$ to XUV irradiation will transform any strongly or mildly
  reducing atmosphere outgassed from the magma ocean into a marginally oxidizing
  atmosphere consisting of H$_2$O, CO$_2$, and N$_2$. Strongly reducing
  atmospheres convert most of their outgassed water reservoir to H$_2$ (which is
  lost) and CO$_2$ in this self-oxidation process.  \\
  \item Self-oxidized atmospheres, as well as primordially oxidizing
  atmospheres, are not prone to HCN production by lightning. This finding
  contrasts with the view that HCN was a key prebiotic feedstock molecule
  \citep{MillerUrey1959}.  Instead, lightning in CO$_2$-rich atmospheres
  produces copious amounts of NO. This NO will dissolve in the oceans and
  transform to nitrate that carries fixed nitrogen available as feedstock to an
  early biosphere. \\
  \item Our results therefore highlight serpentinization (producing H$_2$ and
  metallic FeNi alloy that drive prebiotic carbon fixation) and lightning as key
  planetary processes that converted oxidizing atmospheric components to fixed
  carbon and nitrogen ripe for the abiogenesis process. As an alternative
  to the serpentinization scenario, C may be fixed from CO$_2$ using metallic
  iron from meteorites and volcanic particles as catalyst
  \citep{Peters+etal2023}.\\
  \item The external delivery of fixed nitrogen by interplanetary dust particles
  to an emerging biosphere was likely comparable to the lightning fixation,
  under reasonable assumptions about the early flux of such dust particles.
  Interplanetary dust particles nevertheless display a large range of nitrogen
  concentration and the actual delivered amount would depend on their source
  region (inner or outer Solar System). \\
  \item Extrapolating to rocky planets around other stars, our proposed
  self-oxidation mechanism suggests that the atmospheres of cool rocky planets
  are likely dominated by CO$_2$. The discovery of rocky exoplanets with CO-rich
  or H$_2$-rich atmospheres would be surprising, given the efficiency of
  self-oxidation by hydrogen escape through XUV irradiation.
\end{enumerate}
These conclusions are necessarily based on a number of model assumptions, which
come with a varying degree of certainty. We chose to study two end-members of
the chemical model: one where the molecules CH$_4$ and NH$_3$ are assumed to be
irreversibly destroyed by photolysis and another where those molecules are
assumed to survive. We could then demonstrate that self-oxidation by loss of H
is efficient under both these assumptions. Two salient limitations of our
work are the assumption of chemical equilibrium and our simplified approach to
the XUV irradiation of the thermosphere where we assume that all molecules are
atomized. Future studies should therefore include realistic photochemical
reactions
\citep{Line+etal2011,Kopparapu+etal2012,Zahnle+etal2020,Moses+etal2022} both
during cooling and during XUV-driven mass loss. The energy-limited approach to
XUV mass loss comes with significant uncertainties in both the components
included in the model as well as in the efficiency factor and the planetary
cross section for XUV absorption. We nevertheless showed that loss of H on a
time-scale of a few ten million years is a robust outcome of our calculations
relatively irrespectively of the parameter choice for the XUV mass loss.

We additionally neglected the role of asteroid impacts in generating temporary
reducing atmospheric conditions \citep{Hashimoto+etal2007, SchaeferFegley2017,
Zahnle+etal2020, Itcovitz+etal2022, Gaillard+etal2022, Pearce+etal2022,
Wogan+etal2023}. While such impacts can certainly occur, we would question an
origin-of-life pathway that starts from HCN in a temporary reducing atmosphere
and later completely changes its biochemistry to mimic chemical reactions in a
serpentinization-related hydrothermal system, the proposed metabolism for LUCA
\citep{Weiss+etal2016}, after the loss of the reducing atmosphere and
re-outgassing of an oxidizing atmosphere by volcanism.  We nevertheless
recognize that the origin of life is sufficiently complex that both planetary
processes and external forcing (i.e., impacts of small and large objects) likely
played significant roles.

\begin{acknowledgements}

The authors would like to thank the referees Edwin Kite and Kevin Zahnle for
their many constructive comments to the original manuscript. A.J.\ acknowledges
funding from the European Research Foundation (ERC Consolidator Grant
724687-PLANETESYS), the Knut and Alice Wallenberg Foundation (Wallenberg Scholar
Grant 2019.0442), the Swedish Research Council (Project Grant 2018-04867), the
Danish National Research Foundation (DNRF Chair Grant DNRF159), the G\"oran
Gustafsson Foundation and the Carlsberg Foundation (Semper Ardens: Advance grant
FIRSTATMO). E.C.\ thanks the University of Texas System for a STARs award.
E.v.K.\ acknowledges support from the Villum Young Investigator Grant (project
no.\ 53024). H.J.H.\ acknowledges partial financial support from The Fund of the
Walter Gyllenberg Foundation.

\end{acknowledgements}

%\bibliography{bibliography}
%\bibliographystyle{ar-style1.bst}

\appendix

\section{XUV mass loss model}

We build our XUV mass loss model on the method described in
\cite{Johansen+etal2023b}. However, we have done some important modifications to
the method that allow for a more realistic calculation of the XUV flux from the
host star and the loss of atoms and molecules heavier than H. We present here
the updated model and show tests of how sensitive the mass loss results
presented in the main paper are to the application of a diffusion limit to the
escape rate, to the partitioning of escape energy between H and other atoms, and
finally on our assumption that all molecules in the thermosphere are photolyzed
to atoms. These comparisons are presented in Figure
\ref{f:atmosphere_loss_variation}.

\subsection{Energy-limited escape}

We use the energy-limited escape equation from \cite{Salz+etal2016},
\begin{equation}
  \dot{M}_{\rm el} = \frac{3 \beta^2 \eta F_{\rm XUV}}{4 K G
  \label{eq:MdotEL}
\rho_{\rm p}} \, .
\end{equation}
Here, $\dot{M}_{\rm el}$ is the mass loss rate, $\beta \equiv R_{\rm XUV}/R_{\rm
p}$ quantifies the XUV absorption radius $R_{\rm XUV}$ relative to the planetary
radius $R_{\rm p}$, $\eta$ is the efficiency of the mass loss relative to the
energy-limited case (equivalent to setting $\eta=1$), $F_{\rm XUV} = L_{\rm
XUV}/(4 \pi r^2) $ denotes the flux of XUV radiation at the planet's distance
$r$ to its host star of XUV luminosity $L_{\rm XUV}$, $K$ is factor of order
unity that quantifies the weakened gravity at the escape radius, $G$ is the
gravitational constant and $\rho_{\rm p}$ is the internal density of the planet.
The scaling laws provided in \cite{Salz+etal2016} yield for an Earth-mass planet
in an Earth-like orbit $\beta \approx 2$ (with a very weak dependency on $F_{\rm
XUV}$), $\eta \approx 0.38$ and $K \approx 1$. The simulations of
\cite{Salz+etal2016} nevertheless do not model planets quite down to Earth's
mass, so we take a conservative estimate of $\eta=0.3$ instead of applying their
efficiency scaling law outside of its calibrated range. Other authors have
applied $\eta$ in the range between 0.1 and 0.4
\citep{Lopez+etal2012,Erkaev+etal2014,LugerBarnes2015,Wordsworth+etal2018}.
Regarding the XUV absorption radius, \cite{LugerBarnes2015} used $\beta=1$ while
\cite{Erkaev+etal2014} found for a Mars-like atmosphere $\beta$ in the range
between 3.2 and 4.5; hence energy-limited escape calculations come with high
uncertainties already from the choice of parameters. We ignore also any
potential contributions to mass loss through FUV absorption by H$_2$O molecules
\citep{Sekiya+etal1981}.

We base the XUV luminosity of the host star on the calculations presented in
\cite{Johnstone+etal2021}. These authors consider the distribution of rotation
rates of young stars, with some solar-mass stars initially rotating at up to 100
times the rotation frequency of the modern Sun. Assuming an empirical scaling
between rotation rate and the X-ray luminosity, \cite{Johnstone+etal2021}
calculate the X-ray luminosity as a function of age for stars of different
masses and for different positions along the cumulative distribution of initial
rotation rates. Using an additional empirical scaling law for the connection
between EUV flux (10-92 nm) and X-ray flux (0.1-10 nm), the authors calculate
the full XUV luminosity as a function of stellar age. We downloaded the relevant
XUV data files from the repository provided by the authors of
\cite{Johnstone+etal2021}.

We then follow the approach of \cite{Tian2015} and consider the energy-limited
mass loss to apply not only to the atmosphere's H component, but to the full
atmospheric constituents. This yields the mass loss of component $i$ as
\begin{equation}
  \dot{N}_i = \frac{x_i X_i \dot{N}_{\rm H}^{\star}}{\sum_i X_i
  (\mu_i/\mu_{\rm H})}
  \label{eq:Ndoti}
\end{equation}
Here $N_i$ is the number of component $i$ in the atmosphere, $x_i \leq 1$ is the
fractionation factor (see below), $X_i$ is the mixing ratio of that component in
the thermosphere, $\mu_i$ is the molecular weight of component $i$, and
$\dot{N}_{\rm H}^{\star}$ is the equivalent number loss of H atoms that
corresponds to the energy limited mass loss,
\begin{equation}
  \dot{N}_{\rm H}^{\star} = \dot{M}_{\rm el} / \mu_{\rm H} \, .
  \label{eq:Ndots}
\end{equation}
In the absence of fractionation ($x_i=1$), the sum over $\dot{N}_i \mu_i$ in
equation (\ref{eq:Ndoti}) thus gives $\dot{N}_{\rm H}^{\star} \mu_{\rm
H}=\dot{M}_{\rm el}$ as required for the energy-limited mass loss.

\subsection{Fractionation}

As by far the lightest species, H atoms constitute the primary escape flux.
Heavier atoms and molecules are dragged out with these escaping H atoms as long
as the H flux is sufficiently high. However, as the H flux decreases with time,
the heavier species are subjected to decreased amount of drag and hence obtain
speeds far slower than the H atoms. This leads to a species fractionation that
will change the bulk composition of the atmosphere. Following
\cite{ZahnleKasting1986}, \cite{Hunten+etal1987} and \cite{Erkaev+etal2014}, we
consider atmospheric escape of two dominant escaping species. We assume these
species to be H and O.  This yields the fractionation factor of O
\begin{equation}
  x_{\rm O} = 1 - \frac{b_{\rm H,O} g X_{\rm H}}{k_{\rm B} T F_{\rm H}}
  (\mu_{\rm O}-\mu_{\rm H}) \, .
  \label{eq:xO}
\end{equation}
Here, $b_{{\rm H,O}}$ is the binary diffusion coefficient of O through H, $g$ is
the gravitational acceleration, $X_{\rm H}$ is the molar mixing ratio of H in
the thermosphere, $k_{\rm B}$ is the Boltzmann constant, $T$ is the temperature
and $F_{\rm H}$ is the escape flux of H atoms. We relate the H flux to the
escape rate of H through $F_{\rm H} = \dot{N}_{\rm H}/(3 \pi R_{\rm p}^2)$,
under the assumption that the flux is concentrated towards the day side of the
planet \citep{Erkaev+etal2014}. The fractionation factor is then used to
calculate the escape rate of O through $\dot{N}_{\rm O} = x_{\rm O} X_{\rm O}
\dot{N}$, where $\dot{N}=\sum_i \dot{N}_i$. The escape rate drops to zero when
$x_{\rm O} \le 0$.

We treat C and N as trace species in our nominal model where all molecules
are destroyed in the thermosphere (we discuss the validity of this assumption
below), using the analytical escape model derived in \cite{ZahnleKasting1986}
and further discussed in \cite{Tian+etal2018}. As a trace species, C obtains a
fractionation factor of
\begin{eqnarray}
  x_{\rm C} &=& [1 - g (\mu_{\rm C}-\mu_{\rm H}) b_{\rm H,C}/(F_{\rm H}
  k_{\rm B} T) + (1-x_{\rm O}) f_{\rm O} b_{\rm H,C}/b_{\rm H,O} +
  \nonumber \\
  && x_{\rm O} f_{\rm O} b_{\rm H,C}/b_{\rm O,C}]/(1+f_{\rm O} b_{\rm
  H,C}/b_{\rm O,C}) \, .
  \label{eq:xC}
\end{eqnarray}
Here $f_{\rm O} = X_{\rm O}/X_{\rm H}$. We use a similar expression for the
fractionation factor of N. Treating C this way as a trace species is not
entirely justified, but we tested different approaches to the escape flux of C
and found no difference in how R/O changes with time and only small differences
in the temporal evolution of C/H. The relevant values for
the binary diffusion coefficients were taken from \cite{ZahnleKasting1986} where
provided, except for $b_{\rm O,C}$ which was estimated by
\cite{Erkaev+etal2014}. Similar to \cite{Erkaev+etal2014}, we assume $b_{\rm
H,C}=b_{\rm H,N}=b_{\rm H,O}$.

\subsection{Water in the thermosphere}

Additional assumptions must be made about the water fraction in the
thermosphere. The release of H by photolysis of water is essential to models of
H$_2$O and CO$_2$ loss from Venus and Mars
\citep{WordsworthPierrehumbert2013,Erkaev+etal2014} as well from exoplanets
exposed to high levels of XUV irradiation from the host star
\citep{Tian2015,LugerBarnes2015,Teixeira+etal2024}.

We assume that water vapor is present in the thermosphere at the same
mixing ratio as at the homopause, which is the height over the surface where
energy transport transitions from convective to radiative. We employ a simple
two-component atmosphere model where the energy transport is convective from the
surface up to the photosphere (which we thus associate with the homopause),
above which the remaining atmosphere is optically thin to radiation
\citep{WordsworthPierrehumbert2013}. We set the temperature at the top of the
atmosphere equal to the skin temperature $T_{\rm skin}=T_{\rm eff}/2^{1/4}$
\citep{WordsworthPierrehumbert2013}. Here $T_{\rm eff}$ is the effective
temperature given by
\begin{equation}
  T_{\rm eff} = [(1-A) L_\star / (16 \pi \sigma_{\rm SB})]^{1/4} \, ,
\end{equation}
where $A$ is the albedo, $L_{\star}$ is the bolometric luminosity of the star
and $\sigma_{\rm SB}$ is the Stefan-Boltzmann constant. We calculate the
pressure at the photosphere by integrating the hydrostatic equilibrium from the
top of the atmosphere (where the pressure is zero) down to an optical depth of
$\tau=2/3$. We assume a grey molecular opacity power law $\kappa = \kappa_0
(P/P_0)^\beta$ with $\beta=1$ and an opacity level of $\kappa_{0,{\rm
CO_2}}=0.01\,{\rm kg\,m^{-2}}$ for CO$_2$ and $\kappa_{0,{\rm H_2O}}=1.0\,{\rm
kg\,m^{-2}}$ at $P_0 = 1\,{\rm atm}$ and $\beta=1$ for CO$_2$. This high value
for the gray molecular opacity of H$_2$O was found in \cite{Johansen+etal2023b}
to give a good match to the transition to run-away greenhouse effect in a pure
water atmosphere \citep[we refer to][for detailed
discussions]{Johansen+etal2023b}. For models including CH$_4$ and NH$_3$, we set
for simplicity $\kappa_{0,{\rm CH_4}}=\kappa_{0,{\rm NH_3}}=\kappa_{0,{\rm
CO_2}}$. All other molecules are assumed to carry zero opacity.
\begin{figure*}
  \begin{center}
    \includegraphics[width=0.8\linewidth]{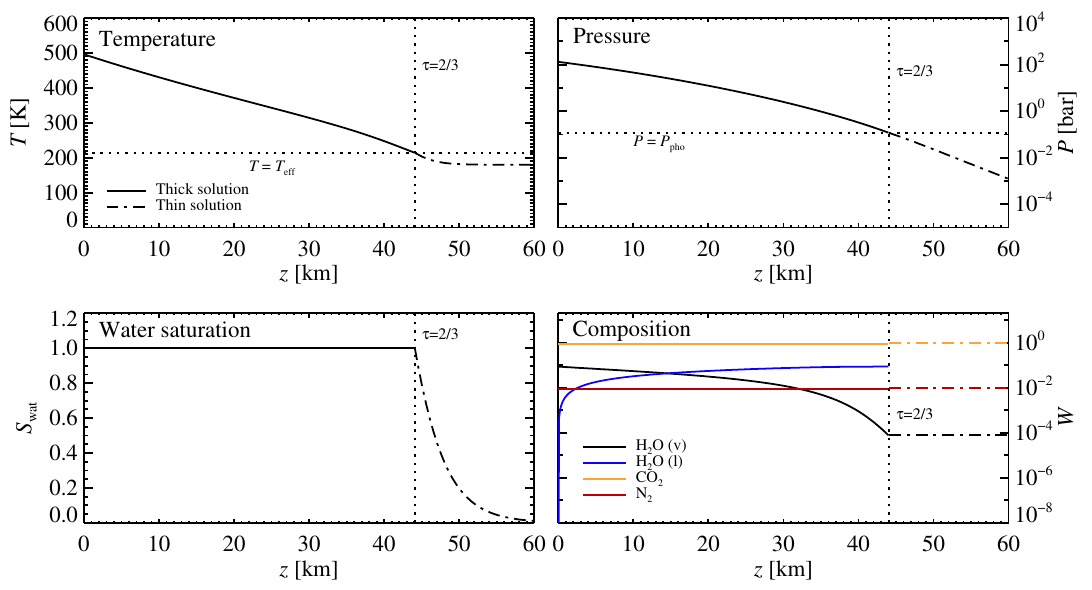}
  \end{center}
  \caption{The temperature, pressure, water saturation level and
  mass-weighted compositions for our equilibrium convective-radiative solution
  to an atmosphere containing only H$_2$O, CO$_2$ and N$_2$. The radiative layer
  above the photosphere is assumed to inherit the water vapor mixing ratio from
  the photosphere where the temperature is equal to the effective temperature;
  no water droplets are allowed to be transported there due to the absence of
  convective motion. The water vapor level above the photosphere is found by an
  iterative procedure to include self-consistently the contribution of the water
  vapor to the opacity.}
  \label{f:equilibrium_atmosphere}
\end{figure*}

The pressure at the photosphere becomes now
\begin{equation}
  P_{\rm pho}=[(1+\beta) P_0^\beta g (2/3)/\kappa_0]^{1/(1+\beta)} \, ,
  \label{eq:Ppho}
\end{equation}
where $\kappa_0$ is the mass-weighted average of the molecular opacities at a
pressure level of $P_0$. All non-condensible molecular species are assumed to
mix efficiently from the photosphere to the thermosphere. We set the mixing
ratio of H$_2$O vapor in the thermosphere to
\begin{equation}
  X_{\rm H_2O}^{\dagger} = {\rm min}(P_{\rm sat,eff}/P_{\rm pho},X_{\rm H_2O})
  \, ,
\end{equation}
where $P_{\rm sat,eff}$ is the saturated vapor pressure of water at the
effective temperature, $X_{\rm H_2O}$ is the mixing ratio of H$_2$O in chemical
equilibrium and $X_{\rm H_2O}^{\dagger}$ is the actual mixing ratio at the
photosphere. We assume that turbulent motion maintains a constant mixing
ratio $X_{\rm H_2O}=X_{\rm H_2O}^\dagger$ above the photosphere. Even small
amounts of water vapor can contribute to increasing the opacity in the
photosphere, so equation (\ref{eq:Ppho}) must therefore be solved iteratively.
In the first iteration we calculate the opacity based on the non-condensible
species only and we subsequently plug in the updated value of $X_{\rm
H_2O}^{\dagger}$ until self-consistency is reached between photospheric pressure
and opacity.

The thermal structure of the convective region is strictly not necessary to
calculate the atmospheric loss, since knowledge of the photospheric pressure and
temperature is sufficient to estimate the mixing ratio of H$_2$O vapor above the
photosphere. We nevertheless present a calculation of the convective region here
to illustrate its connection to the photosphere and the optically thin regions.
We integrate hydrostatic equilibrium from the surface up until reaching the
photospheric pressure. The temperature gradient follows a mixed, wet adibat
\citep{WordsworthPierrehumbert2013}. The surface pressure is given by $P_{\rm
sur} = g M_{\rm atm}/(4 \pi R_{\rm p}^2)$. The wet adiabat approach ensures that
water vapor is always present at its local saturation. We remove any excess of
water vapor at the surface of the planet to form an ocean.  Water vapor that
condenses to clouds within the atmosphere are kept as tiny droplets that have a
very large molecular weight, so that the clouds contribute to the density of the
atmosphere but not to the pressure. All atmospheric constituents are assumed to
follow an ideal gas law with parameterized fits to the specific heat capacity
taken from the NIST Chemistry
WebBook\footnote{\url{https://webbook.nist.gov/chemistry/}}.

We show in Figure \ref{f:equilibrium_atmosphere} the temperature, pressure,
water saturation and atmosphere composition as a function of height for an
oxidizing atmosphere with $M_{\rm CO_2}=M_{\rm H_2O}=10^{-4}\,M_{\rm E}$ and
$M_{\rm N_2}=10^{-6}\,M_{\rm E}$. We set the stellar luminosity to 70\% of the
Sun's modern value and an albedo of $A=0.5$. The optically thin solution
$T(\tau) = [(3/4) \tau + 1/2]^{1/4} T_{\rm eff}$ is shown with a dashed line; we
emphasize that the convective solution is obtained by requiring continuity in
$P$ and $T$ at $\tau=2/3$. This self-consistency is obtained by iterating over
the surface temperature. We do not allow water droplets to be transported into
the radiative region above the photosphere. At conditions of nearly constant
temperature and falling pressure, the optically thin regions therefore become
undersaturated in water vapor. The water mixing ratio in the thermosphere is
very low, at the level of $\sim$$10^{-4}$. Hence water is efficiently
cold-trapped in our models and largely protected from escaping. This motivates
us to not pursue more complex models of water mass loss here; we refer to
\cite{ZahnleKasting2023} for more realistic models of water mass loss from a
strongly irradiated early Venus.

\begin{figure}
  \begin{center}
    \includegraphics[width=0.8\linewidth]{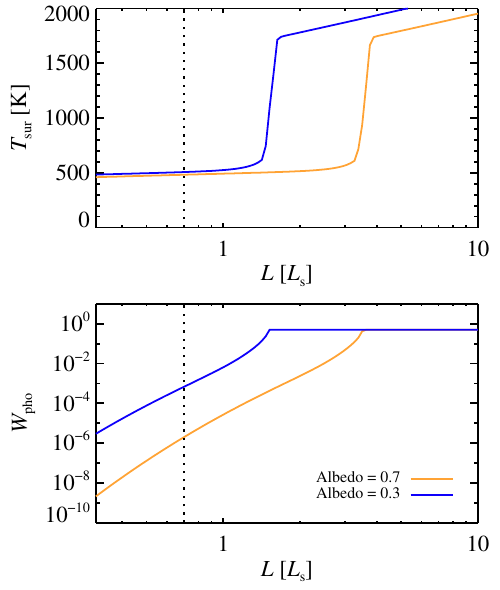}
  \end{center}
  \caption{The surface temperature and water vapor mixing ratio at the
  photosphere of a planet at 1 AU distance from the host star as a function of
  the luminosity of the star. The likely luminosity of the young Sun is
  indicated with a dotted line. Two different albedo values are chosen to
  represent the modern Venus and Earth end-members. The transition to run-away
  greenhouse effect (and a moist photosphere) depends strongly on the albedo of
  the planet but happens in both cases at insolation levels in excesss of modern
  Earth's.}
  \label{f:surface_temperature}
\end{figure}

In Figure \ref{f:surface_temperature} we show how the surface temperature
and photospheric water mixing ratio depend on the stellar luminosity for a
planet at 1 AU distance from its host star. We analyse two different values of
the albedo ($A=0.7$ and $A=0.3$) to represent end-member values for modern Venus
and modern Earth. The transition to run-away greenhouse effect (and the
associated wet photosphere) requires insolations in significant excess of
Earth's current level \citep{Kasting1988}. Since we focus in this work on
planets under terrestrial insolation or lower, Figure
\ref{f:surface_temperature} demonstrates that planets in our model have an
efficient cold trap that prevents significant loss of water.

\subsection{Energy redistribution}

The heavier species O, C and N fractionate significantly relative to H through
equations (\ref{eq:xO}) and (\ref{eq:xC}). This fractionation implies that the
actual number flux from equation (\ref{eq:Ndoti}) will not sum up to the
energy-limited $\dot{N}_{\rm H}^\star$ from equation (\ref{eq:Ndots}). An energy
conserving analytical expression exists for a two-component system with H and a
second species \citep{Tian2015}. However, since we use multiple species, we
solve equation (\ref{eq:Ndoti}) by increasing the value of $\dot{N}_{\rm H}$
until equation (\ref{eq:MdotEL}) is fulfilled under self-consistent calculation
of the fractionation factors $x_i$ (which depend on the hydrogen flux). This
energy partitioning between H and other species leads to the surprising behavior
that even a very low hydrogen mixing ratio $X_{\rm H} \ll 1$ in the thermosphere
can lift an atmosphere of high mean molecular weight at the energy-limited rate
\citep{Tian2015}. We therefore scale the energy-limited loss efficiency by the
mixing ratio of H in the thermosphere, setting the effective mass loss
efficiency $\eta' = \eta X_{\rm H}$, to recover the limit of extremely
inefficient mass loss from a CO$_2$-dominated atmosphere \citep{Tian2009}.

\subsection{Diffusion limit}

We also consider that the supply of H to the thermosphere may be diffusion
limited. The diffusive flux of H to the thermosphere through a background of
mean molecular weight $\bar{\mu}$ can be calculated from
\begin{equation}
  F_{\rm H}^{\rm (diff)} = \frac{b g (\bar{\mu}-\mu_{\rm H}) X_{\rm H}}{k_{\rm
  B} T} \, .
\end{equation}
Here $b$ is the binary diffusion coefficient and $g$ is the gravitational
acceleration. We use $b=6.5 \times 10^{19} T^{0.7}\,{\rm m^{-1}\,s^{-1}}$
relevant for H diffusing through air \citep{Hunten1973}. This expression
formally describes the maximum H flux that can be extracted from the system when
the heavier background gas does not escape
\citep{Hunten1973,Zahnle+etal1990,WordsworthPierrehumbert2013,LugerBarnes2015}.
However, the expression more generally describes the availability of H in the
thermosphere by diffusion from the lower atmosphere layers \citep[see discussion
in][]{ZahnleCatling2017}. In our models, this availability becomes an issue once
$X_{\rm H} \ll 1$ in the atmosphere. We refer to \cite{ZahnleKasting2023} for
more advanced diffusion models involving multiple diffusing species.
\begin{figure*}
  \begin{center}
    \includegraphics[width=\linewidth]{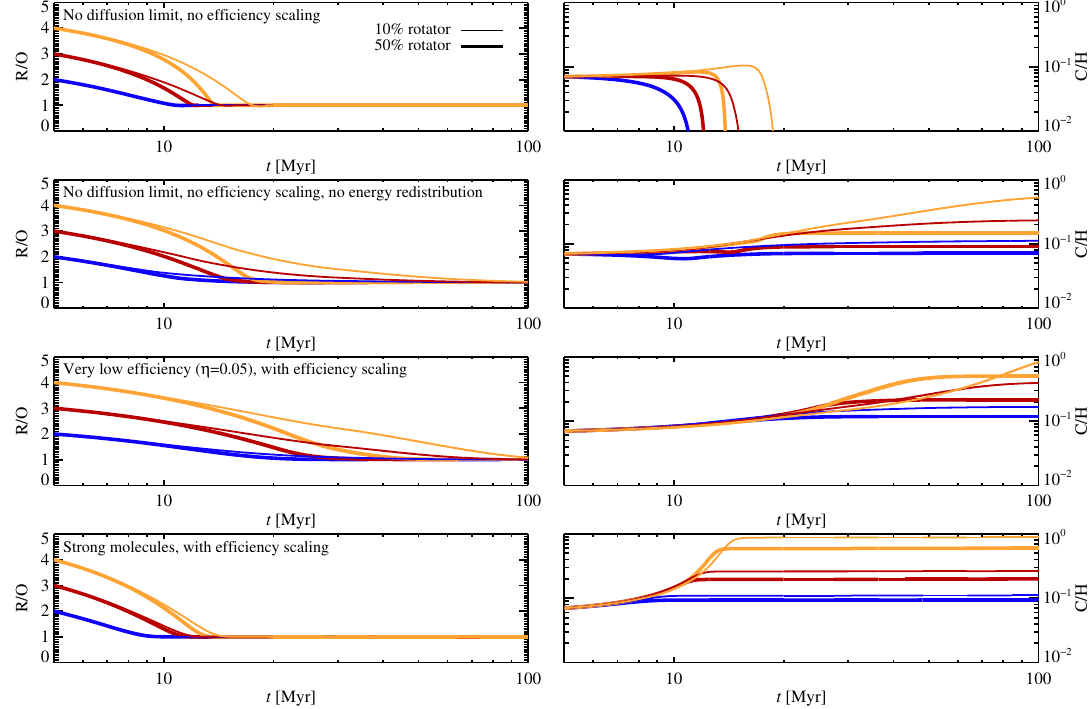}
  \end{center}
  \caption{Variations of the XUV mass loss model for atmospheres without
  CH$_4$ and NH$_3$. The top panel shows results when we do not include the
  diffusion limit or efficiency scaling with $X_{\rm H}$. Mass loss is very
  rapid in that case, blowing off the entire C budget by the redistribution of
  mass-loss energy from H to C, N and O.  This leaves behind an atmosphere with
  R/O=1 and consisting entirely of H$_2$O.  In the second row we ignore energy
  redistribution from non-escaping heavy atoms to H. The C/H ratio rises in that
  case, as the H flux is insufficient to drive off the heavier atoms.  In the
  third row we show the effect of reducing the mass loss efficiency to
  $\eta=0.05$. Self-oxidation takes up to 50 million years for the slowest
  rotator. The bottom plot shows the atmospheric evolution when we assume that
  only H$_2$ is destroyed in the thermosphere. This leads to a very strong
  fractionation in C/H due to maintenance of the heavier CO$_2$ and CO
  molecules.}
  \label{f:atmosphere_loss_variation}
\end{figure*}

\subsection{Variations to mass loss model}

We apply both energy partitioning and either mass-loss efficiency scaled to
$X_{\rm H}$ or the diffusion limit to our nominal calculations presented in the
main paper. We note that those effects work in opposite directions: efficient
partitioning of energy from H to heavier species increases the mass loss rate,
while the efficiency scaling and the diffusion limit act to decrease the mass
loss rate. In our nominal mass loss model, we also assume that the XUV
irradiation is intense enough during the first 100 million years of stellar
evolution that all molecules are efficiently photolysed to atoms in the
thermosphere \citep{Erkaev+etal2014}. However, this assumption may not be
realistic for strongly bound molecules such as CO$_2$ and CO \citep{Tian2009}.
We therefore run also a set of simulations where only H$_2$ is atomized.
Given the lack of data on binary diffusion coefficients of a trace species
diffusing through CO$_2$ and CO in the tables provided in
\cite{ZahnleKasting1986}, we use for simplicity equation (\ref{eq:xO}) as
fractionation factor for all the molecular species. We show the results of
varying some model parameters in Figure \ref{f:atmosphere_loss_variation}.
These calculations are all performed for the model without CH$_4$ and NH$_3$,
since we can then solve the chemical equilibrium system much faster without
using FastChem.

Overall, Figure \ref{f:atmosphere_loss_variation} tells a cautious tale that the
results of XUV mass loss models depend strongly on the included physics and
chemistry as well as a set of relatively uncertain input parameters. But the
figure also reveals that self-oxidation within a few 10 Myr is a robust outcome
of XUV mass loss, independently of the model details.

\section{The bombardment history of Earth}

Planet formation theory combined with the abundances of highly siderophile
elements in the Earth's mantle can be used to shed light on Earth's accretion
history after the main formation phase and the moon-forming giant impact. The
moon-forming giant impact must have been a truly cataclysmic event that melted
a large fraction of the mantle and would have destroyed any putative
life-forms that had evolved in the quiescent period between the end of the main
accretion and the impact. Depending on assumptions on impactor mass and the
equilibration efficiency between the core of the impactor and the mantle of
proto-Earth, the timing of the impactor has been inferred to anywhere between 25
Myr and 150 Myr after the formation of the Sun
\citep{Yin+etal2002,ConnellyBizzarro2016,KleineWalker2017,Johansen+etal2023b}.

The presence of highly siderophile elements of chondritic relative proportions
in the Earth's mantle requires an implantation of 0.5-0.8\% mass into a solid
mantle, which at this time no longer allowed the sinking of metal and
siderophile elements into the core \citep{Walker2009,Marty2012}.
\cite{Brasser+etal2016} divided the collision history of Earth after the
moon-forming giant impact into two phases: the {\it late veneer} that delivered
highly siderophile elements into the Earth's mantle before crust formation 4.42
Gyr ago (approximately 130 Myr after Solar System formation) and {\it late
accretion} that denotes bombardment of Earth after the crust had solidified. The
early timing and deep penetration of the late veneer makes this bombardment
phase irrelevant for the delivery of molecules carrying fixed N to the early
biosphere. Late accretion, on the other hand, would have implanted the Earth's
crust and atmosphere with objects ranging in size from IDPs to 100-km-scale
asteroids. \cite{Brasser+etal2016} used $N$-body simulations to infer a likely
remnant planetesimal mass reservoir of $M_{\rm LA}$$\sim$$10^{-3}\,M_{\rm E}$
after the terrestrial planets had fully formed. Assuming an impactor N abundance
similar to ordinary chondrites and a late accretion time-scale of $\tau_{\rm
LA}$$\sim$$100\,{\rm Myr}$, the rate of fixed N delivery becomes
\begin{equation}
  r_{\rm N,LA} = 8 \times 10^{26}\,{\rm s^{-1}}\,f_{\rm fix} \left(
  \frac{M_{\rm LA}}{10^{-3}\,M_{\rm E}} \right) \left( \frac{\tau_{\rm
  LA}}{100\,{\rm Myr}} \right) \left( \frac{C_{\rm N}}{10\,{\rm ppm}}\right)
\end{equation}
This delivery rate is comparable to the lightning fixation rate on early Earth
derived in Section \ref{s:NOyield}. The relevant value of $f_{\rm fix}$, the
fraction of the asteroid-delivered nitrogen that is not converted to N$_2$ in
the impact, is nevertheless highly uncertain for the late accretion phase,
particularly if asteroidal impacts were hot enough to destroy all the organics
\citep{Anders1989} and convert fixed N into inert N$_2$. The high fraction of
free metal in ordinary chondrite and enstatite chondrites, representing
terrestrial neighbourhood bodies that likely dominated the late accretion flux,
would nevertheless give very reducing conditions in the impact. Such conditions
allow for production of significant amounts of NH$_3$ from very large impactors
in the Vesta-Ceres size range \citep{Zahnle+etal2020}.

\end{document}